\documentclass[10pt,journal,compsoc]{IEEEtran}
\ifCLASSOPTIONcompsoc
  \usepackage[nocompress]{cite}
\else
  \usepackage{cite}
\fi
\ifCLASSINFOpdf
\else
\fi
\usepackage{amsmath}
\usepackage{amsfonts}
\usepackage{amssymb}

\usepackage{url}
\usepackage{graphicx}
\usepackage{multirow}
\usepackage{multicol}

\usepackage[hidelinks]{hyperref}

\usepackage{color}
\usepackage{ulem}

\newcommand{\HL}[1]{\textcolor{black}{#1}}
\newcommand{\ZZH}[1]{\textcolor{black}{#1}}
\newcommand{\NZH}[1]{\textcolor{black}{#1}}
\newcommand{\red}[1]{\textcolor{black}{#1}}
\newcommand{\redd}[1]{\textcolor{black}{#1}}

\begin{document}
%
% paper title
% Titles are generally capitalized except for words such as a, an, and, as,
% at, but, by, for, in, nor, of, on, or, the, to and up, which are usually
% not capitalized unless they are the first or last word of the title.
% Linebreaks \\ can be used within to get better formatting as desired.
% Do not put math or special symbols in the title.
% \title{Motion Retargeting via Body-Parted Level Attention}
\title{Pose-aware Attention Network for Flexible Motion Retargeting by Body Part}

%
% author names and IEEE memberships
% note positions of commas and nonbreaking spaces ( ~ ) LaTeX will not break
% a structure at a ~ so this keeps an author's name from being broken across
% two lines.
% use \thanks{} to gain access to the first footnote area
% a separate \thanks must be used for each paragraph as LaTeX2e's \thanks
% was not built to handle multiple paragraphs
%
%
%\IEEEcompsocitemizethanks is a special \thanks that produces the bulleted
% lists the Computer Society journals use for "first footnote" author
% affiliations. Use \IEEEcompsocthanksitem which works much like \item
% for each affiliation group. When not in compsoc mode,
% \IEEEcompsocitemizethanks becomes like \thanks and
% \IEEEcompsocthanksitem becomes a line break with idention. This
% facilitates dual compilation, although admittedly the differences in the
% desired content of \author between the different types of papers makes a
% one-size-fits-all approach a daunting prospect. For instance, compsoc 
% journal papers have the author affiliations above the "Manuscript
% received ..."  text while in non-compsoc journals this is reversed. Sigh.

% \author{Lei~Hu,~\IEEEmembership{Member,~IEEE,}
%         Zihao~Zhang,~\IEEEmembership{Fellow,~OSA,}
%         Chongyang~Zhong,~\IEEEmembership{Life~Fellow,~IEEE,}
%         Boyuan~Jiang,~\IEEEmembership{Member,~IEEE,}
%         Shihong~Xia,~\IEEEmembership{Member,~IEEE,}
 \author{Lei~Hu$^{\dag}$,
        Zihao~Zhang$^{\dag}$,
        Chongyang~Zhong,
        Boyuan~Jiang,
        Shihong~Xia$^{*}$,
    
        % <-this % stops a space
\IEEEcompsocitemizethanks{
% M. Shell was with the Department
% of Electrical and Computer Engineering, Georgia Institute of Technology, Atlanta,
% GA, 30332.\protect\\
% note need leading \protect in front of \\ to get a newline within \thanks as
% \\ is fragile and will error, could use \hfil\break instead.
% E-mail: see http://www.michaelshell.org/contact.html
\IEEEcompsocthanksitem
$\dag$~Equal contributions \protect\\
\IEEEcompsocthanksitem
*~ Corresponding author\protect\\
\IEEEcompsocthanksitem
Lei Hu, Chongyang Zhong, Boyuan Jiang and Shihong Xia are with Institute of Computing Technology, Chinese Academy of Sciences, Beijing 100190, China, and also with the University of Chinese Academy of Sciences, Beijing 100190, China.\protect\\
E-mail: hulei19z, zhongchongyang, jiangboyuan20s, xsh@ict.ac.cn
\IEEEcompsocthanksitem Zihao Zhang is with Institute of Computing Technology, Chineses Academy of Sciences, Beijing 100190, China.\protect\\
E-mail: zhangzihao@ict.ac.cn}% <-this % stops an unwanted space

% \thanks{Code~:
% \href{https://github.com/hlcdyy/pan-motion-retargeting}{https://github.com/hlcdyy/pan-motion-retargeting}}
}

% note the % following the last \IEEEmembership and also \thanks - 
% these prevent an unwanted space from occurring between the last author name
% and the end of the author line. i.e., if you had this:
% 
% \author{....lastname \thanks{...} \thanks{...} }
%                     ^------------^------------^----Do not want these spaces!
%
% a space would be appended to the last name and could cause every name on that
% line to be shifted left slightly. This is one of those "LaTeX things". For
% instance, "\textbf{A} \textbf{B}" will typeset as "A B" not "AB". To get
% "AB" then you have to do: "\textbf{A}\textbf{B}"
% \thanks is no different in this regard, so shield the last } of each \thanks
% that ends a line with a % and do not let a space in before the next \thanks.
% Spaces after \IEEEmembership other than the last one are OK (and needed) as
% you are supposed to have spaces between the names. For what it is worth,
% this is a minor point as most people would not even notice if the said evil
% space somehow managed to creep in.

% The paper headers
% \markboth{Journal of \LaTeX\ Class Files,~Vol.~14, No.~8, August~2015}%
\markboth{Journal of \LaTeX\ Class Files}
{Shell \MakeLowercase{\textit{et al.}}: Bare Demo of IEEEtran.cls for Computer Society Journals}
% The only time the second header will appear is for the odd numbered pages
% after the title page when using the twoside option.
% 
% *** Note that you probably will NOT want to include the author's ***
% *** name in the headers of peer review papers.                   ***
% You can use \ifCLASSOPTIONpeerreview for conditional compilation here if
% you desire.

% The publisher's ID mark at the bottom of the page is less important with
% Computer Society journal papers as those publications place the marks
% outside of the main text columns and, therefore, unlike regular IEEE
% journals, the available text space is not reduced by their presence.
% If you want to put a publisher's ID mark on the page you can do it like
% this:
%\IEEEpubid{0000--0000/00\$00.00~\copyright~2015 IEEE}
% or like this to get the Computer Society new two part style.
%\IEEEpubid{\makebox[\columnwidth]{\hfill 0000--0000/00/\$00.00~\copyright~2015 IEEE}%
%\hspace{\columnsep}\makebox[\columnwidth]{Published by the IEEE Computer Society\hfill}}
% Remember, if you use this you must call \IEEEpubidadjcol in the second
% column for its text to clear the IEEEpubid mark (Computer Society jorunal
% papers don't need this extra clearance.)

% use for special paper notices
%\IEEEspecialpapernotice{(Invited Paper)}

% for Computer Society papers, we must declare the abstract and index terms
% PRIOR to the title within the \IEEEtitleabstractindextext IEEEtran
% command as these need to go into the title area created by \maketitle.
% As a general rule, do not put math, special symbols or citations
% in the abstract or keywords.
\IEEEtitleabstractindextext{%
\begin{abstract}

\ZZH{Motion retargeting is a fundamental problem in computer graphics and computer vision. Existing approaches usually have many strict requirements, such as the source-target skeletons needing to have the same number of joints or share the same topology. To tackle this problem, we note that skeletons with different structure may have some common body parts despite the differences in joint numbers. Following this observation, we propose a novel, flexible motion retargeting framework. The key idea of our method is to regard the body part as the basic retargeting unit rather than directly retargeting the whole body motion. To enhance the spatial modeling capability of the motion encoder, we introduce a pose-aware attention network (PAN) in the motion encoding phase. The PAN is pose-aware since it can dynamically predict the joint weights within each body part based on the input pose, and then construct a shared latent space for each body part by feature pooling. Extensive experiments show that our approach can generate better motion retargeting results both qualitatively and quantitatively than state-of-the-art methods. Moreover, we also show that our framework can generate reasonable results even for a more challenging retargeting scenario, like retargeting between bipedal and quadrupedal skeletons because of the body part retargeting strategy and PAN. Our code is publicly available\thanks{\href{https://github.com/hlcdyy/pan-motion-retargeting}{$^1$https://github.com/hlcdyy/pan-motion-retargeting}}$^1$.}

\end{abstract}

% Note that keywords are not normally used for peerreview papers.
\begin{IEEEkeywords}
Deep Learning, Motion Processing, Motion retargeting
\end{IEEEkeywords}}

% make the title area
\maketitle

% To allow for easy dual compilation without having to reenter the
% abstract/keywords data, the \IEEEtitleabstractindextext text will
% not be used in maketitle, but will appear (i.e., to be "transported")
% here as \IEEEdisplaynontitleabstractindextext when the compsoc 
% or transmag modes are not selected <OR> if conference mode is selected 
% - because all conference papers position the abstract like regular
% papers do.
\IEEEdisplaynontitleabstractindextext
% \IEEEdisplaynontitleabstractindextext has no effect when using
% compsoc or transmag under a non-conference mode.

% For peer review papers, you can put extra information on the cover
% page as needed:
% \ifCLASSOPTIONpeerreview
% \begin{center} \bfseries EDICS Category: 3-BBND \end{center}
% \fi
%
% For peerreview papers, this IEEEtran command inserts a page break and
% creates the second title. It will be ignored for other modes.
\IEEEpeerreviewmaketitle

\IEEEraisesectionheading{\section{Introduction}\label{sec:introduction}}

\IEEEPARstart{A}rticulated motion data plays a crucial role in \ZZH{computer animation, }virtual reality and the game industry\ZZH{, since most of the virtual characters are driven by the articulated skeletons.} \ZZH{To get the motion data, current methods are mainly two-fold. The first type is to capture human motions using a motion capture system, the other one is to create animations by artists using key-frame technology. However, both methods require a major expenditure of time and effort, \HL{and the obtained motion data cannot be directly applied to each other between skeletons because of the differences in structure.} Motion retargeting, as one of the motion-reusing technology, has been regarded as a promising way that enables the transfer of a character's motion to another skeleton. \NZH{
It is widely used as a motion pre-processing tool since it can integrate various motion datasets~\cite{loper2014mosh} for neural network training.}}
% .Recently, with the development of deep learning technology, motion retargeting is widely used as a }motion pre-processing tool since it enables the integration of various motion datasets for subsequent training. 
In physical simulation and control~\cite{kim2022human, rempe2020contact, ye2022neural3points}, motion retargeting also plays an important role in transferring the input signals from the real environment into the simulation setting.
\begin{figure}[ht]
    \centering
    \includegraphics[width=\linewidth]{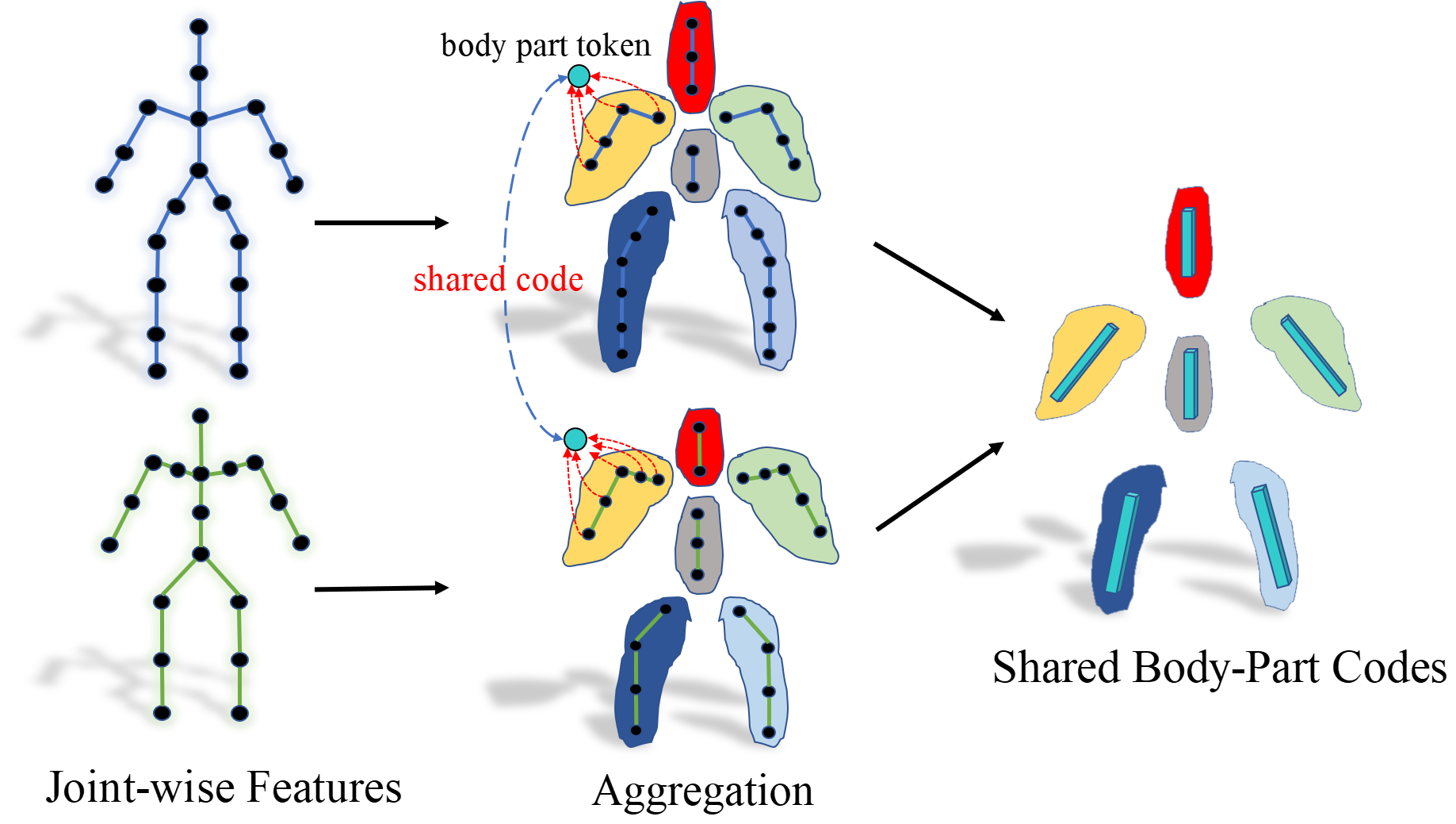}
    \caption{Feature Pooling at body part level. We pool the joint-wise features (black dots) into body part tokens (cyan dots) to build the shared latent codes in terms of body parts.}
    \label{fig:body_part_modeling}
\end{figure}

\ZZH{Though motion retargeting is a long-standing problem that has been studied for decades, there are still two main issues that restrict the large-scale application of this technology. First, motion retargeting essentially requires flexible source-target correspondence. However, we observe that previous motion retargeting methods~\cite{villegas2018neural, lim2019pmnet, villegas2021contact} have some strict requirements, such as the source-target skeletons needing to have the same number of joints or share the same topology. These requirements will reduce the flexibility of the motion retargeting model and limit the usage scenarios.}
% set-up convenience is relaced by automation
Another issue is the balance between the accuracy and automation of motion retargeting. Traditional works~\cite{gleicher1998retargetting, kwang2000line, lee1999hierarchical} regard motion retargeting as a space-time optimization problem and employ the inverse kinematic technology to precisely satisfy the user-given constraints. However, users need to design the energy functions manually, making the whole process semi-automated. \HL{ Recent works~\cite{villegas2018neural, lim2019pmnet, aberman2020skeleton, villegas2021contact} achieve automatic motion retargeting using the deep learning methods. However, the retargeting accuracy is still limited due to the lack of spatial modeling of articulated motion.}

\HL{To tackle the above issues, it first needs to define a flexible correspondence between the source and target skeletons. Therefore,}
\ZZH{we propose to treat the body part as the basic retargeting unit to extract shared latent codes cross structure (see Figure~\ref{fig:body_part_modeling}). Our key observation supporting this strategy is that skeletons with different structure still have some common body parts sharing the same semantic meaning. Using the body part as the retargeting unit not only helps us improve the flexibility of retargeting, but also gives the neural network a geometric prior. 
\HL{To enhance the spatial modeling capability of the model}, we further introduce a novel pose-aware attention network (PAN). Inspired by PFNN~\cite{holden2017phase} and its follow-up works~\cite{zhang2018mode, starke2019neural, zhong2022spatio}, we find that the dynamic motion modeling based on the state/pose is beneficial for generating high-quality motions. Therefore, we use the proposed PAN to dynamically predict the weights of each joint for feature blending and pooling. Moreover, the dynamic spatial modeling meets our intuition that the contribution of a fixed joint to its corresponding body part varies with the body part’s motion.}

% To improve the stability and accuracy of our motion retargeting method, we propose to treat the body part as the basic retargeting unit and introduce a novel pose-aware attention network. Our key observation supporting the body part strategy is that different skeletons still have some common body parts sharing the same semantic meaning. Moreover, we believe that subdividing the whole articulated skeleton into several body parts can give the neural network a geometric prior and therefore improve the stability and accuracy of the retargeting results.
% To improve the correspondence flexibility of the motion retargeting method, we introduce the trainable parameters called "body part tokens" and compute the dot-product attention weights together with the hidden features of inner-part joints. Inspired by previous work~\cite{holden2017phase, zhang2018mode, starke2019neural, zhong2022spatio}, we find that the dynamic motion modeling based on the the state/pose is beneficial for generating high-quality motions. The dynamic weight is therefore predicted to empower the network focus differently based on the input.} % The importance of each joint in the corresponding part may be different and the dynamic motion modeling~\cite{holden2017phase, zhang2018mode, starke2019neural, zhong2022spatio} based on the the state/pose is beneficial for generating high-quality motions.

Given an articulated motion \ZZH{of} the source skeleton, we first process the motion frame-by-frame using our PAN to extract the spatial features at the body part level. \HL{Specifically, we introduce a trainable} parameter for each body part called "body part token" (see cyan dots in Figure~\ref{fig:body_part_modeling}) and compute the dot-product attention weights together with the hidden features of the inner-part joints. Through several attention layers, the joint-wise features will be blended into the "body part tokens" by weighted summation. For feature pooling, we keep the parameters of the body parts, i.e. "body part tokens" and discard the hidden features of joints in the last layer, so that these "body part tokens" can be shared among skeletons with different structure. Then, we further compress the "body part tokens" along the temporal dimension by 1D convolution to extract the shared motion code. Finally, we combine the shared motion code and the deep representation of the target skeleton offsets to generate the retargeted motion by the motion decoder of the target structure. 

% Specifically, We introduce learnable parameters called "body part tokens" and compute the dot-product attention weights together with the hidden features of inner-part joints. Though several attention layers, the joint-wise Value features will be aggregated to the "body part tokens" by weighted summation. For the purpose of pooling, we keep the parameters of the body parts and discard the joint features, so that these body part-wise features can be shared among skeletons with different structure. Then, we further compress the body part-wise features along the temporal dimension by 1D convolution for extracting the shared motion code. Finally, we combine the shared motion code and the deep representation of the target skeleton offsets to generate the retargeted motion by the motion decoder of target structure. 

Since paired motion data is hard to acquire in this task, we train our architecture in an unsupervised manner and use a motion discriminator for each articulated structure to ensure the retargeted motion falls into the corresponding motion manifold. 
Experiments on Mixamo~\cite{mixamo}, \red{Human3.6M~\cite{ionescu2013human3}, }lafan1~\cite{harvey2020robust}, and quadruped~\cite{zhang2018mode} datasets show that our method can achieve state-of-the-art performance.

Our approach can effectively address the issues mentioned above, enabling automatic, accurate, and flexible retargeting. The main contributions of this work can be summarised as follows:

\ZZH{1. We propose a novel pose-aware attention network (PAN) that can dynamically extract the spatial features of motion, which is beneficial for improving the accuracy of automatic retargeting. }

\ZZH{2. We propose a powerful motion retargeting framework that uses body parts as retargeting units, improving the flexibility in processing different source-target structure pairs.}

\ZZH{3. Through extensive experiments, we show that our method can generate high-quality results for human motion retargeting task. Because of our body-part retargeting strategy and pose-aware attention network, we find that our method can even fulfill the retargeting task between bipedal and quadrupedal skeletons. To the best of our knowledge, we are the first to solve such motion retargeting problems without manual effort.}

\section{Related work}
In this section, we will review related research on motion retargeting, motion processing with body parts, and dynamical motion modeling methods that are similar to our thinking. 
\subsection{Motion Retargeting}
One of the earliest motion retargeting methods was proposed by Gleicher~\cite{gleicher1998retargetting}, which regards retargeting as a spacetime constraint problem. As most of the constraints in retargeting are kinematic related, inverse kinematics solver~\cite{lee1999hierarchical,kwang2000line, hecker2008real} is widely adopted for this task. Besides, there is some literature~\cite{popovic1999physically,tak2005physically} that takes dynamics into consideration for preserving essential physical properties of the motion during retargeting. When the source and target skeletons differ greatly, the key-framing technique plays an important role in many works~\cite{hecker2008real, yamane2010animating,seol2013creature}. Yamana et al.~\cite{yamane2010animating} combine the key-frame selection with a GPLVM-based model to learn the static and dynamic mapping between the source and target poses for anthropomorphic characters' retargeting. Yeongho Seol et al.~\cite{seol2013creature} ask the users to manually specify several features representing the source motion and build a paired data library through key-framing. However, most traditional motion retargeting approaches rely on hand-crafted constraints and iterative solvers, making the workflow semi-automatic and tedious. 

Recently, with the fast development of deep learning and articulated motion datasets, end-to-end motion retargeting methods have achieved remarkable success. Jang et al.~\cite{uk2020variational} use convolutional encoder-decoder architecture to model the source motion and limb ratios for retargeting between skeletons with different bone lengths. 
Villegas et al.~\cite{villegas2018neural} train an RNN-based encoder-decoder network to alternate inverse kinematics module. Aberman et al.~\cite{aberman2020skeleton} propose the skeleton-aware convolution and pooling operations to extract a primal skeleton for handling topologically equivalent retargeting. In order to reduce interpenetration and preserve self-contacts, Villegas et al.~\cite{villegas2021contact} combine iterative optimization and geometry-conditioned RNN to achieve real-time retargeting between different mesh geometries. Though the deep learning-based methods achieve automatic motion retargeting, the lack of spatial modeling of the skeletal structure makes the retargeting accuracy limited. Skeleton-aware network~\cite{aberman2020skeleton} can handle different structure to some extent, but the flexibility of their correspondence strategy is limited because all its operations are based on the neighborhoods in the kinematic chain. These operations do not work well for constructing correspondence in some cases, such as retargeting between quadrupeds and bipeds.
% In addition to 3D motion retargeting, Aberman et al.~\cite{aberman2019learning} propose a 2D retargeting approach video to video using separate encoders to resolve the view, skeleton, and motion in the video.

% For the motion retargeting from human to non-humanoid character, because of the variance in skeleton morphology, it is more difficult for the users to define the correspondence between these two skeletons. Hecker et al.~\cite{hecker2008real} propose a generalized motion and use it to convert the source motion to the target motion through inverse kinematics. An anthropomorphic motion retargeting method is proposed by Yamana et al.~\cite{yamane2010animating}. Given the selected key frames of the human pose, the method obtains the retargeted motion through a GPLVM-based method. Yeongho Seol et al.~\cite{seol2013creature} ask the users to manually specify several features which could represent the input motion and use linear mapping for retargeting. Similarly, this method also requires manual selection of key frames and tags when building a paired data library. In addition to the motion retargeting, Helge Rhodin et al.~\cite{rhodin2015generalizing} proposed a gesture-based motion control method to drive the animal skeleton in real-time.Still, their approach has to be combined with sensors and VR devices.

\subsection{Motion Processing with Body Parts}
\begin{figure*}[!t]
    \centering
    \includegraphics[width=\linewidth]{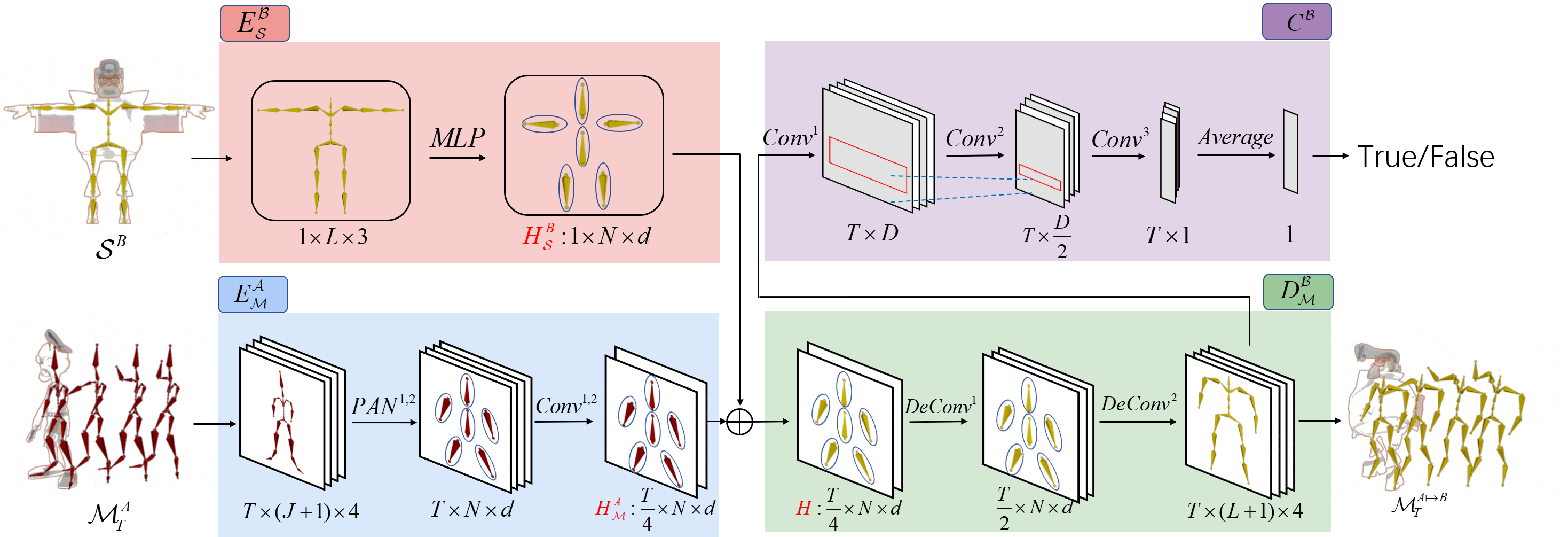}
    \caption{The overall architecture of our motion retargeting framework. The transparent meshes are overlaid on the first frame of source motion and the canonical pose of the target skeleton to highlight the differences between skeleton $A$ and $B$. The motion code $H_\mathcal{M}^A$ produced by $E_\mathcal{M}^\mathcal{A}$ will be modified by skeleton code $H_\mathcal{S}^B$ in an additive manner, i.e, $H = H_\mathcal{M}^A + H_\mathcal{S}^B$ ($H_\mathcal{S}^B$ will first be replicated along the temporal axis.)}
    \label{fig:overview_framework}
\end{figure*}
\HL{Subdividing the whole body motion into several partial movements is widely used in motion splicing and style transfer. Based on the source of the partial movements, we can divide the body part-based methods into two categories. 1. The movement of each body part comes from different motions. 2. The movement of each body part comes from a single motion. Our body part-based retargeting method falls into the second category.}

For the first category, the difficulty of splicing lies in choosing the appropriate combination of partial motions, since the movements of the individual body parts are usually coupled with each other. To tackle this problem, traditional methods use segmentation, clustering~\cite{jang2008enriching}, and example-based techniques~\cite{heck2006splicing, ma2010modeling} to measure the similarity of partial motions to ensure the naturalness of the splicing motion. However, these splices are often performed in the original space or linear embedding space reduced by PCA, thus limiting the splicing to analogous motions. Thanks to the non-linear modeling of deep neural networks, recent methods have tried to splice the motion features of different body parts in the hidden space. Ye et al.~\cite{ye2022neural3points} use three points of the VR device as input and predict the movement of the upper and lower body respectively, and then splice them together. Motion Puzzle~\cite{jang2022motion} transfers the style of each body part from different motion clips to the content motion by graph convolutional network. \red{Lee et al.~\cite{lee2022learning} propose a novel part assembler layer for splicing the part motions from the different creatures. This assembler can search for the spatial alignment among body parts after the temporal alignment. }These works show that it is a good choice to combine the motion features of different body parts in deep motion space. 

There is no ambiguity in the splicing phase for the second category because all the partial movements come from the same motion clip. However, if the target skeleton is different from the source motion skeleton, we still need to modify the motion features of each body part to match the target skeleton.  Abdul-Massih et al.~\cite{abdul2017motion} decompose motion style into a set of features present in distinct groups of body parts and use optimization methods to solve the retargeting between skeletons with different morphologies. However, this approach requires manual feature design and spatial-temporal alignment, making the whole process semi-automatic. Liao et al.~\cite{liao2022skeleton} propose a skeleton-free pose transfer network to automatically learn the skinning weights and transformation in terms of deformation parts. But, it can only process a single pose rather than motion. We regard the body parts as retargeting units and employ deep neural networks to automatically construct the shared latent spaces between skeletons with different structure, but have common body parts.

\subsection{Dynamical Motion Modeling}
\HL{In motion retargeting, we need to encode the source motion features, which is actually a spatio-temporal modeling process. \red{Dynamic modeling of articulated motion has been included in many studies of various tasks. In this part, we will briefly review some of the }methods related to our methodological ideas.} Dynamical motion modeling is crucial for many applications, including style transfer, motion synthesis, prediction, etc. We divide dynamic motion modeling into two categories, one for dynamically changing the network's weights such as PFNN~\cite{holden2017phase}, and the other is dynamically changing the feature vectors such as Transfomer~\cite{vaswani2017attention}.

For dynamically changing the control weights, Xia~\cite{xia2015realtime} proposed an online mixture-of-autoregressive model for real-time motion style interpolation and control by blending the parameters of distinctive styles. PFNN~\cite{holden2017phase} has achieved remarkable success in data-driven motion synthesis thanks to its phase-functioned model. It works by generating the weights of a regression network at each frame as a function of the phase, thus enabling smooth transitions between different motion states with good stability and producing high-quality motion. Follow-up works~\cite{zhang2018mode, starke2019neural, 2020Local, starke2021neural} employ a gating network to automatically learn the blending coefficients of different experts which allows the network to cluster the poses based on different states or phases. \red{The graph convolutional neural network is also a common tool for modeling human body motion, but the adjacency matrix of the vanilla network is fixed. To make the graph connections more flexible to portray the relationships between joints, Lei Shi et al.~\cite{shi2019two} add a learnable matrix and a data-dependent graph to change the adjacency matrix based on the input data.} Recently, Zhong et al.~\cite{zhong2022spatio} introduce the gating adjacency matrix into the graph convolution network for motion prediction, and the core idea is also to dynamically change the graph convolutional weights to improve the expression of the model. Although the mixture-of-experts model has excellent motion modeling capabilities, the size of the model increases proportionally with the number of experts. 

The idea of the Transformer~\cite{vaswani2017attention} is similar to the mixture-of-experts model, except that the dynamic modeling is achieved by generating attention weights through computing the dot products of the query with key vectors. The advantage of this model is that the network size can be controlled to stack more layers. This architecture is currently widely used for pose estimation~\cite{lin2021end, zheng20213d, zhang2022mixste}, \red{action-gesture recognition~\cite{shi2020decoupled}} and motion synthesis~\cite{petrovich2021action}. \red{In motion style transfer, Jang et al.~\cite{jang2022motion} use attention network transfers the locally semantic style features into the decoded content features. In this work, we employ the self-attention mechanism in source motion encoding process }and incorporate the body part strategy to achieve pose-aware spatial feature extraction for motion retargeting. The proposed pose-aware attention mechanism can dynamically generate the attention weights based on the input poses to aggregate the joint-wise features at the body part level. Because of the dynamic spatial modeling, we are able to achieve more accurate motion retargeting.

\subsection{Other Related Works}
There are some loosely related problems including mesh deformation transfer and 2D motion retargeting. Deformation transfer~\cite{sumner2004deformation, baran2009semantic, celikcan2015example, gao2018automatic} aims to adapt the deformation from a source mesh model to another mesh. These methods often require building dense correspondences between the meshes. Recently, Gao~\cite{gao2018automatic} proposes a method that can transfer the deformation between two unpaired mesh models without defining the correspondences. The cycle consistency loss used in their work is similar to ours, except that we impose consistency constraints on both the latent space and the original rotation space and focus on the articulated skeleton. 

There are also some works~\cite{ren2021flow, aberman2019learning} that bypass the explicit 3D motion representation to directly implement 2D motion retargeting, but these often incorporate viewpoint and texture generation. In contrast, we mainly focus on 3D motion retargeting in this work. 

\section{Data representation and Overview}\label{sec:data_rep}
\ZZH{In this section, we will first describe the representation of articulated motion data, skeleton representation, and the symbols used in our paper. Then, we will further introduce the overview architecture of our method.}

\subsection{Data representation}

% Before discussing our retargeting framework in detail, we first describe the representation of articulated motion data, skeleton representation, and the symbols used in our paper. 

We denote an articulated motion of length $T$ as $\mathcal{M}_T=[M_1, M_2, ..., M_T]$ and suppose \HL{that the skeleton $A$ belonging to structure} $\mathcal{A}$ has $J$ joints. The motion attributes used in our representation include local joint rotations of each joint represented by unit quaternions denoted as $Q^{T\times J\times 4}$, global motion velocity $V^{T\times 1\times 3}$ of the root in x,y, and z directions and additional angular velocity $R^{T\times 1\times 1}$ representing the rotation velocity around the axis perpendicular to the ground (y-axis in our setup). We concatenate the \ZZH{velocity $V$ and angular velocity }$R$ in the last dimension to represent a new combined velocity vector $\bar{V}^{T\times 1\times4}$, and the motion $\mathcal{M}_T$ can be written as $\mathcal{M}_T = [Q, \bar{V}]\in \mathbb{R}^{T\times (J+1)\times 4}$ if we regard the root velocity vector $\bar{V}$ as an additional "joint". In articulated skeleton representation, \HL{the skeleton topology and bone lengths} are usually represented by a set of offsets $S\in \mathbb{R}^{J\times 3}$,  the offset of each joint in the skeleton is the 3D vector relative to its parent's coordinate frame in the kinematic chain. The task of motion retargeting is adapting a motion $\mathcal{M}_T^A$ from structure $\mathcal{A}$ with offsets $\mathcal{S}^A\in \mathbb{R}^{J\times 3}$ to skeleton $B\in \mathcal{B}$ with offsets $\mathcal{S}^B\in \mathbb{R}^{L\times 3}$. The retargeted motion is $\mathcal{M}_T^{A\mapsto B} \in \mathbb{R}^{T\times (L+1)\times 4}$ where the joint number $L$ may not be equal to $J$, but has the same time length $T$. 

\HL{In humanoid motion retargeting, we divide the whole body into $N$=6 body parts based on the main limbs, which are the head, the spine, the left/right arms, and the left/right legs. \red{Specially, we attribute the velocity of root joint to all body parts as special "joint" because its important role in distinguishing the whole-body motion types.} To achieve flexible source-target correspondence, we will only construct the common body part motion spaces of the source and target skeletons in practice. For example, when we retarget the motion from the structure of a normal person to the skeleton of a disabled person with only one arm, we construct only $N$=5 shared body parts. In addition to this, we allow semantic-level correspondence when facing more complex retargeting tasks such as bipeds to quadrupeds, which will be discussed in Sec~\ref{sec:quadruped}.}

\subsection{Overview}

\HL{Figure~\ref{fig:overview_framework} shows the overall architecture of our approach. \red{Our pipeline is similar to that of the motion Puzzle which is designed for motion style transfer at the body part level. However, our task differs from style transfer in that we take the target skeleton representation rather than the target motions as input.} As shown on the left side of the figure, the input of the framework is the source motion $\mathcal{M}_T^A$ performed by skeleton $A\in \mathcal{A}$ and targeted skeleton offsets $\mathcal{S}^B$. The output of our framework is the motion $\mathcal{M}_T^{A\mapsto B}$ that is performed by skeleton $B\in \mathcal{B}$.  Our architecture can be divided into four modules, namely, skeleton encoder $E_{\mathcal{S}}^\mathcal{B}$, motion encoder $E_{\mathcal{M}}^\mathcal{A}$, motion decoder $D_{\mathcal{M}}^\mathcal{B}$ and motion discriminator $C^\mathcal{B}$. The skeleton encoder $E_{\mathcal{S}}^\mathcal{B}$ receives the target skeleton offsets as input and encodes them by a multi-layer perception to generate the skeleton code $H_\mathcal{S}^B$. The target skeleton code is composed of the hidden features of $N$ body parts, each with $d$ dimensions. 
\ZZH{The motion encoder $E_{\mathcal{M}}^\mathcal{A}$ is mainly in charge of mapping the source motion $\mathcal{M}_T^A$ into the source motion code $H_\mathcal{M}^A$ . It includes the pose-aware attention network (PAN) for spatial modeling and the convolution layers for temporal compression. }
% The source motion $\mathcal{M}_T^A$ will pass through the motion encoder $E_{\mathcal{M}}^\mathcal{A}$, which includes the PAN for spatial modeling and the Conv layers for temporal compression. The generated source motion code $H_{M}$ has the common body parts and hidden dimension $d$ with the skeleton code $H_\mathcal{S}^B$. 
The motion code $H_\mathcal{M}^A$ and the skeleton code $H_\mathcal{S}^B$ will be fused in an additive manner and pass through the motion decoder $D_{\mathcal{M}}^\mathcal{B}$ to generate the retargeted motion $\mathcal{M}_T^{A\mapsto B}$. 
Since our networks are trained in an unsupervised manner, we build a discriminator ($C^\mathcal{B}$ in Figure~\ref{fig:overview_framework}) for each structure to ensure the retargeted motion falls onto the correspondence motion manifold.}

As stated above, in order to enhance the spatial modeling capability of the model, we introduce the pose-aware attention network (PAN) in the motion encoder $E_{\mathcal{M}}^\mathcal{A}$ to extract the source motion features in terms of body parts. In the following sections, we will first elaborate on this network. 
\begin{figure*}[ht]
    \centering
    \includegraphics[width=\linewidth]{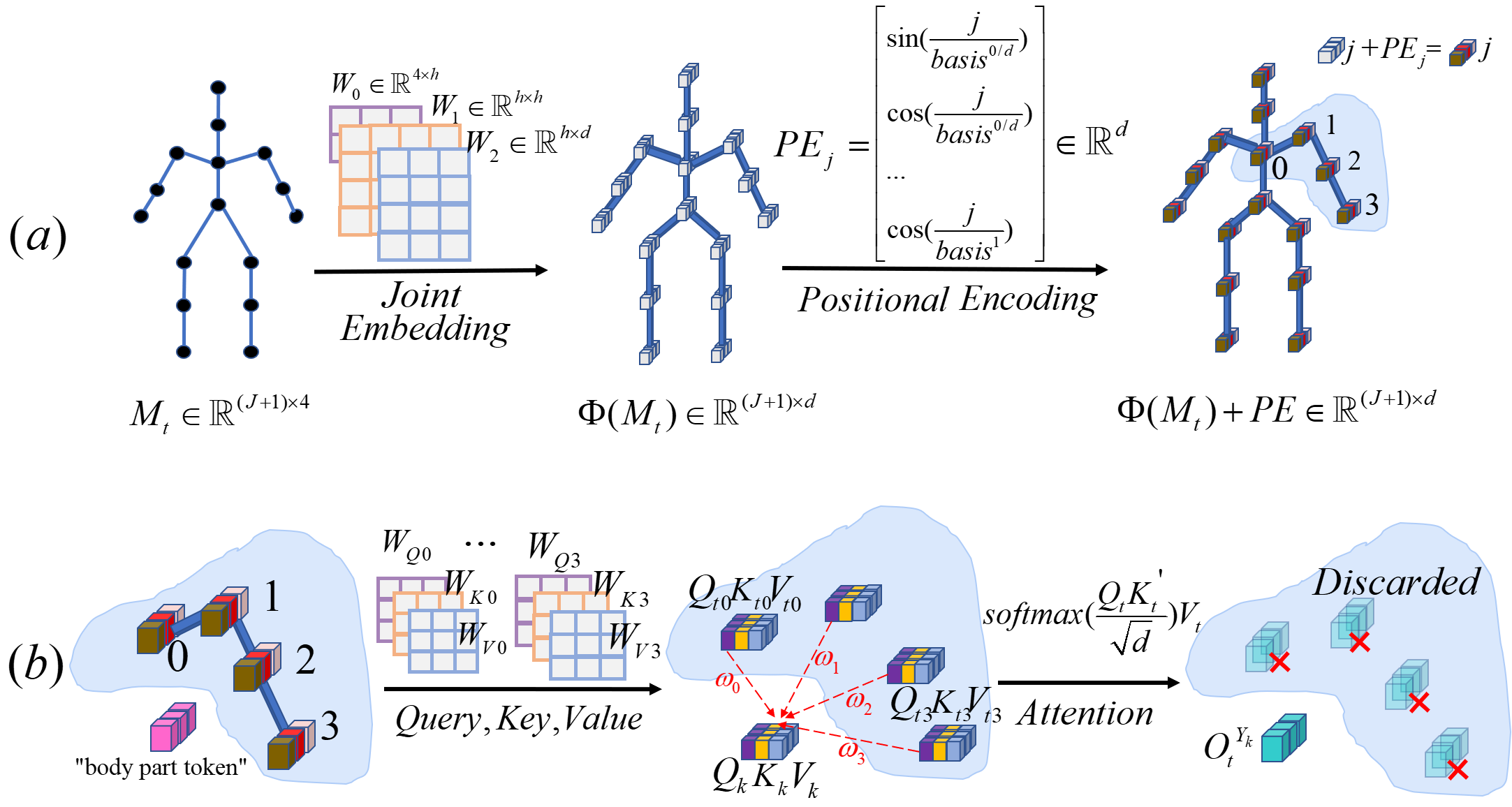}
    \caption{(a) Joint embedding and positional encoding. The raw joint features will first be embedded by MLP and then modified by positional encodings. (b) Pose-aware attention at the body part level. We use the arm part as a sample to demonstrate the attention process. In the last layer of the attention, the output features $O_t^{Y_k}$ corresponding to the "body part token" will be retained, while other nodes will be discarded.}
    \label{fig:pos_encoding}
\end{figure*}

\section{Pose-Aware Attention Network}\label{sec:attention_networks}

In this section, we will introduce the most important component of \HL{the motion encoder}, i.e, the pose-aware attention network(PAN). \ZZH{We will first describe the joint embedding and positional encoding, and finally illustrate our pose-aware attention mechanism.} 

The choice of using attention networks for spatial feature extraction is based on the idea that the importance of each joint in the kinematic chain is different and varies dynamically with the pose (The visualization shown in Sec~\ref{sec:experiments} will confirm this assumption). This prompts us to employ a neural network to automatically learn the attention weights, aggregating the motion characteristics at the body part level. Transformer~\cite{vaswani2017attention} achieves remarkable success in natural language processing due to the capacity of the attention mechanism which could automatically depict the association of separate words in a sentence. Inspired by that, we try to model the spatial relationships between joints using the proposed pose-aware attention network.

\subsection{Joint Embedding and Positional Encoding}\label{sec:pos_encoding}
\NZH{The PAN will process the motion frame-by-frame. Given the input pose represented as $M_t\in \mathbb{R}^{(J+1)\times 4}$ at the timestep $t$, we need to first map the input to the hidden embedding space through the joint embedding layer, which is shown in Figure~\ref{fig:pos_encoding} (a). This step can be formulated as follows:}
\begin{equation}
    \Phi(M_t; \alpha) = Relu(Relu(M_tW_0+b_0)W_1+b_1)W_2+b_2
    \label{eq:joint_embedding}
\end{equation}
Where the parameters of the network $\alpha$ are defined by $\alpha=\{W_0\in \mathbb{R}^{4\times h}, W_1\in \mathbb{R}^{h\times h}, W_2 \in \mathbb{R}^{h\times d}, b_0 \in \mathbb{R}^{h}, b_1 \in \mathbb{R}^{h}, b_2 \in \mathbb{R}^{d}\}$. Here $h$ is the number of hidden units used in non-linear mapping which is 256 in our implementation and the $d$ is the dimensionality of the joint embedding space. 

\NZH{However, directly using the above joint feature is not feasible since the network lacks the ability to know where each joint is located in the kinematic chain. Inspired by the  positional encoding in the vanilla attention network (Transformer), we use the joint-level positional encoding to label each joint's location in an additive manner. }In our case, we modify the mathematical formulation of positional encodings and use the joint indices in the kinematic chain as the position. The formula is represented as follows: 
\begin{equation}
    \begin{aligned}
        PE_{j, 2i} = \NZH{sin(\frac{j}{basis^{2i/d}})}\\
        PE_{j, 2i+1} = cos(\frac{j}{basis^{2i/d}})
    \end{aligned}
    \label{eq:pos_encoding}
\end{equation}

where $j$ indicates the index of the joint in the kinematic chain, $d$ equals to the dimensionality of the embeddings in equation~\ref{eq:joint_embedding} and $i\in [0,...,d/2]$. The $basis$ which can control the sinusoid's frequency is often set to 10,000 as in most transformer implementations. The joint-level positional encoding is bounded, smooth and dense compared to the one-hot form and the vector dimension of $PE$ is not correlated with the number of skeleton joints, which makes it very suitable for encoding in the case of skeleton structure changes.

After calculating the $PE_j$ for each joint, we further combine the joint embeddings with the positional encodings in an additive manner, which is given by the following equation:
\begin{equation}
    X_{t,j} = \Phi(M_{t,j};\alpha) + PE_j \qquad \forall j \in J
    \label{eq:additive_joint}
\end{equation}
where $M_{t,j}$ represents the input feature of joint $j$ in pose $M_{t}$. 
Through the non-linear joint embedding and positional encoding, the hidden representation $X_t$ contains information about each joint rotation and location in the kinematic chain.

\subsection{Pose-aware Attention at Body Part Level}

As our retargeting strategy is to use body parts as shared units, we need to integrate and pool joint-wise features into body part-wise codes.
\NZH{Previous literature~\cite{petrovich2021action} proposes to use the "distribution tokens" to pool arbitrary-length motion sequences into one latent space. Inspired by this work, we}

similarly prepend the joint embeddings with learnable tokens and only use the corresponding attention outputs as a way to pool the joint-wise features into body part level. 
Specifically, we introduce a learnable parameter with the same dimension $d$ as the joint embedding for each body part, thus yielding a vector $Y \in \mathbb{R}^{N\times d}$ which we called "body part tokens", where the $N$ represent the body part number we have defined. Figure~\ref{fig:pos_encoding}(b) shows the operations with an arm part as an example. For each timestep $t$, the attention function first takes as input a %concatenated variables
\NZH{combination of the body part tokens and the joint-level features} $Z_t=[Y, X_t] \in \mathbb{R}^{(N+J+1)\times d}$ and outputs the Query vector $Q_t$ and Key-Value pair ($K_t$ and $V_t$). the formulation can be described as follow:
\begin{equation}
    \begin{aligned}
    % Q_t &= Z_t W_Q + b_Q\\
    % K_t &= Z_t W_K + b_K\\
    % V_t &= Z_t W_V + b_V
    Q_t = Z_t W_Q + b_Q \quad K_t = Z_t W_K + b_K \quad V_t = Z_t W_V + b_V
    \end{aligned}
    \label{eq:QKV}
\end{equation}
where $W_{[Q, K, V]}\in \mathbb{R}^{d\times d}$ and $b_{[Q, K, V]}\in \mathbb{R}^d$ are learnable matrix and bias vectors, respectively. 

Since we regard the body part as a retargeting unit, we will use a mask matrix $U \in \mathbb{R}^{(N+J+1)\times (N+J+1)}$ to isolate the body parts %to guarantee 
\NZH{so that }the final body-part features are generated purely by pooling the inner-part joints' features. The mask matrix $U$ is symmetric, where $U_{i,j} = 0$ means that joint $i$ and $j$ are in the same body part and vice verse $U_{i,j} = -\infty$. The final output of the attention is computed as a weighted sum of the values: 
\begin{equation}
    Attention(Q_t, K_t, V_t, U) = softmax(\frac{Q_tK_t^{'}+U}{\sqrt{d}})V_t
    \label{eq:attention_function}
\end{equation}
where $K_t^{'}$ represents the transpose of $K_t$ and $\sqrt{d}$ \NZH{is used for scaling to prevent pushing the softmax function into regions where it has extremely small gradients~\cite{vaswani2017attention}.}
% is used for scaling to prevent pushing the softmax function into regions where it has extremely small gradients~\cite{vaswani2017attention}. 
% We are able to interpret the equation~\ref{eq:attention_function} from the skeleton view, for each query vector $Q_{t, i}$ of joint $i$ at timestep $t$, it calculates the correlation with each joint key $K_{t, j}$ of the same body part(including itself) measured by dot production $Dot(Q_{t, i}, K_{t, j})$, and then weighted sum of value vector $V_t$ according to the dot-product attention. It is worth noting that the Dot-production attention is not symmetric since the $Dot(Q_{t, i}, K_{t, j})$ may different with $Dot(Q_{t, j}, K_{t, i})$. 
The production of Query $Q$ and the transpose of Key $K$ actually depict the relationship of the joints. \NZH{Specifically, the larger the dot-production value of two joints reveals the stronger their relationship.}
% A larger value of the dot product indicates a stronger dependency between these two joints.
This attention operation can be stacked into several layers to form the PAN, in our case we use 2 layers. In the last layer, we discard the inner-part joint values and only retain the "body part token" values (see Figure~\ref{fig:pos_encoding}(b) ). 

% \NZH{Substitute equation~\ref{eq:QKV} in equation~\ref{eq:attention_function}, we can easily get a new formulation $\Psi(Z_t; \beta)$ representing the relationship of input concatenate vector $Z_t$ and PAN parameters $\beta = \{W_Q, W_K, W_V, b_Q, b_K, b_V, U\}$. Since the attention weights $\Psi(Z_t; \beta)$ is dependent on variables $Z_t$, it is also related to the input pose $M_t$ according to equation ~\ref{eq:additive_joint}. Therefore, our attention network is pose-aware and able to dynamically extract the spatial features of the articulated skeleton.}
Substitute equation~\ref{eq:QKV} in equation~\ref{eq:attention_function}, we can easily get a new formulation $\Psi(Z_t; \beta)$, where the parameters are defined by $\beta = \{W_Q, W_K, W_V, b_Q, b_K, b_V, U\}$ (all except $U$ are trainable). Our attention network is pose-aware since the attention weights calculated by $Q_tK_t^{'}$ are dependent on the variable $Z_t$, which is related to the input pose $M_t$ according to equation~\ref{eq:additive_joint}. Therefore, the proposed PAN can dynamically extract the spatial features of the articulated motion.

% All parameters in $\beta$ except $U$ are trainable.
\NZH{As stated above, we only use the output corresponding to the "body part token". Therefore, denote the complete output of the $\Psi(Z_t; \beta)$ as $O_t = [O_t^Y, O_t^X] \in \mathbb{R}^{(N+J+1)\times d}$, we only keep the $O_t^Y \in \mathbb{R}^{N \times d}$ for pooling purpose.}
% Combining the equations~\ref{eq:QKV} and~\ref{eq:attention_function}, we can obtain a new formulation $\Psi(Z_t; \beta)$, where the parameters of attention network $\beta$ are defined by $\beta = \{W_Q, W_K, W_V, b_Q, b_K, b_V, U\}$, and all parameters except $U$ are trainable. 
% Suppose the output matrix of attention network $\Psi(Z_t; \beta)$ is $O_t = [O^Y, O^X] \in \mathbb{R}^{(N+J+1)\times d}$, we only use the outputs $O^Y \in \mathbb{R}^{(N \times d)}$ for body-part feature pooling.

% We can find that the attention weights calculated by $Q_tK_t^{'}$ are dependent on variables $Z_t$, which is inferred from formulation~\ref{eq:additive_joint} thus related to the input pose $M_t$. Therefore, our attention network is pose-aware and able to dynamically extract the spatial features of the articulated skeleton. In Sec~\ref{sec:experiments} we will demonstrate that pose-aware attention is beneficial for improving the accuracy of motion retargeting. 

\section{Architecture Modules and Training Process}
In this section, we will describe in detail the modules of our architecture including the motion encoder, skeleton encoder, motion decoder, and discriminator. Then we will show the training/testing process and the loss functions we used. 
\subsection{Motion Encoder}
\NZH{The motion encoder is designed to extract the source motion code $H_\mathcal{M}^A$ from spatio-temporal dimension. The process of motion encoding is shown in the blue part ($E_\mathcal{M}^\mathcal{A}$) of Figure~\ref{fig:overview_framework}. The motion encoder consists of two blocks, the pose-aware attention PAN$^{i},i\in \{1, 2\}$ and the temporal convolutional $Conv^{i}, i\in \{1, 2\}$. Given a source motion $\mathcal{M}_T^A$, the pose-aware attention network generate the hidden features of $N$ body parts, i.e, $O^Y = [O^{Y_1}, O^{Y_2}, ...,O^{Y_N}], O^{Y_k}\in \mathbb{R}^{T\times d}$. The PAN is mainly in charge of extracting spatial information. As for the temporal compression, we use the multi-level temporal convolution $Conv$, which is proven to be influential in building motion manifold~\cite{holden2015learning, holden2016deep} as well as retargeting~\cite{aberman2020skeleton}. In our setting, the convolution block $Conv$ takes each body part features $O^{Y_k}$ as input, and the temporal modeling process can be described as follows:}
\begin{equation}
    Conv_k = Relu(O^{Y_k} \ast W^{conv}_k + b^{conv}_k) \qquad \forall k\in {1, 2, ...,N}
    \label{eq:conv}
\end{equation}
\NZH{where $\ast$ means the convolution operation, $W^{conv}_k\in \mathbb{R}^{h\times d\times w}$ is the weights matrix with temporal filter width of $w$, and $b^{conv}_k \in \mathbb{R}^{h}$ is the bias with $h$ hidden units in the convolutional layer. The filter width $w$ is set to 15 so that the receptive field of the filter can roughly cover a half second of motion, which is proven to be efficient in work~\cite{holden2016deep}. We set the number of hidden units to 32 for each body part because we experimentally find it can produce good reconstruction and retargeting results. }The stride of all the convolution kernels is set to 2 for temporal compression purpose. Through the $Conv^{i}, i\in \{1, 2\}$ we eventually obtain the shared motion code $H_\mathcal{M}^A\in \mathbb{R}^{\frac{T}{4}\times N\times d}$. 

\subsection{Skeleton Encoder}
The skeleton encoder \NZH{(see the pink part ($E_\mathcal{S}^B$)} of Figure~\ref{fig:overview_framework}) aims to transform the target skeleton offsets into latent codes in terms of body parts. Since the offset vectors, which express the skeleton topology and bone lengths, can be regarded as a single canonical pose, we only spatially encode the offset vectors. Given the target skeleton offsets $S^B$ with $L$ joints, we utilize a multi-layer perceptron (three layers in our work) to map the raw vectors to the skeleton code $H_\mathcal{S}^B \in \mathbb{R}^{1\times N\times d}$. Specifically, for each body part, we concatenate the corresponding joints' offsets to form a vector $S_k\in \mathbb{R}^{3L_k}$, $k\in [1, 2, ..., N]$ and pass through the following equation:
\begin{equation}
    \omega(S_k; \gamma_k) = Relu(Relu(S_kW_{k0}+b_{k0})W_{k1}+b_{k1})W_{k2}+b_{k2}
\end{equation} 
Where the parameters of MLP are defined by $\gamma_k = \{W_{k0}\in \mathbb{R}^{3L_k\times h}, W_{k1}\in \mathbb{R}^{h\times h}, W_{k2}\in \mathbb{R}^{h\times d}, b_{k0}\in \mathbb{R}^h, b_{k1}\in \mathbb{R}^h, b_{k2}\in \mathbb{R}^d\}$. Here $h$ is the number of hidden units which is 64 in our implementation, and $L_k$ is the number of joints in $k$th body part. We stack the latent codes belonging to different body parts to generate the final skeleton code $H_\mathcal{S}^{B} = [\omega(S_1; \gamma_1), ..., \omega(S_N; \gamma_N)]\in \mathbb{R}^{1\times N\times d}$

\subsection{Motion Decoder}
Through the motion encoder $E_\mathcal{M}^\mathcal{A}$ and skeleton encoder $E_\mathcal{S}^\mathcal{B}$, we obtain the source motion code $H_\mathcal{M}^A$ and target skeleton code $H_\mathcal{S}^B$, respectively. To integrate these two codes, we first repeat the target skeleton code $H_\mathcal{S}^B$ along the temporal axis to make it consistent with the shape of $H_\mathcal{M}^A$ and then add it to $H_\mathcal{M}^A$. \red{The reason we directly add them is our motion and skeleton codes are both composed in the form of body parts. Direct summation enables body-part codes to correspond to each other (e.g., arm motion and arm skeleton)} The motion decoder (see green part ($D_\mathcal{M}^\mathcal{B}$) in Figure~\ref{fig:overview_framework}) receives the fused motion code $H$, and generate the retargeted motion $\mathcal{M}_T^{A\mapsto B}$ by $Deconv$ blocks. The $Deconv$ block consists of up-sampling and deconvolution as follows:
\begin{equation}
    Deconv = Relu(\uparrow H \ast W^{dec} + b^{dec})
\end{equation}
Where the weights matrix $W^{dec}\in \mathbb{R}^{h\times Nd\times w}$ and bias vector $b^{dec} \in \mathbb{R}^{h}$ are defined similar with formulation~\ref{eq:conv}. 
\NZH{The difference is that we }integrate all the body-part features to synthesize the retargeted motion of the whole body \red{i.e., the deconvolutional kernel $W^{dec}$ is not body part-wise and will act on the fused features $H$ to consider the relationship between different body parts}. The number of $Deconv$ blocks is 2 and we use the same kernel width and hidden size with $Conv$ blocks (see equation~\ref{eq:conv}), but the stride of the kernel is 1. We set the up-sampling factor as 2 in order to recover the temporal dimension layer by layer\red{, and in the last layer we restore the spatial dimension of the target motion, i.e.$Nd\rightarrow (L+1)\times4$. It should be
noted that the attention weights computed in the motion encoder will not be used for decoding}. In the last layer, the $Relu$ activation will not be used and the convolution operation will synthesize the retargeted motion performed by skeleton $B$ with the same sequence length $T$. 

\subsection{Motion Discriminator}
The retargeted motion generated by the decoder needs to fall onto the motion manifold of the target structure. Therefore, we build a motion discriminator for each skeletal structure as shown in the purple part ($C^\mathcal{B}$) of Figure~\ref{fig:overview_framework}. The discriminator $C^\mathcal{B}$ consists of several convolution layers and receives the motion $\mathcal{M}_T^{A\mapsto B}$ as input. Each layer of $C^\mathcal{B}$ aims to progressively compress the feature size from $4(L+1)$ to $D/2$, here the $D$ is set to 256. %(the convolutional operation is similar with equation~\ref{eq:conv}, but not body part-related) and finally output a one-dimension feature which is the range in (0, 1) using sigmoid function as follow:
\NZH{The convolutional operation is similar with equation~\ref{eq:conv}, but not body part-related. The final output of the motion discriminator is a one-dimension feature in the range of (0, 1) by using the sigmoid function:}
\begin{equation}
    \sigma(x) = \frac{1}{1+e^{-x}}
\end{equation}
The average score indicates the naturalness of the retargeted sequence. We train the discriminator and motion generator alternatively to encourage the motion decoder to synthesize natural-looking motions, the detailed process is demonstrated in section~\ref{sec:training_process}

\subsection{Training and Testing Process}\label{sec:training_process}
One of the main difficulties our method faces is acquiring paired motion data in real-world scenarios. Even if the same actor repeatedly performs the same motion clip, there will still be slight differences. We follow the training strategy of SAN~\cite{aberman2020skeleton}, which uses cyclic adversarial learning to train the encoder-decoder pairs of different structure for the purpose of self-supervision. This strategy can overcome the data deficiency problem.
% Previous deep learning-based motion retargeting works~\cite{villegas2018neural, lim2019pmnet, aberman2020skeleton} train the networks in an unsupervised manner to avoid overfitting. In fact, it is extremely difficult for us to acquire paired motion data in real-world scenarios because even if the same actor repeatedly performs the same motion clip, there will still be differences. Therefore, we also use the unsupervised/self-supervised training strategy. 
Specifically, we take the motion encoder, skeleton encoder, and motion decoder together as a set called generator $G = \{E_\mathcal{M}, E_\mathcal{S}, D_\mathcal{M}\}$, we alternatively update the parameters between generator $G$ and discriminator $C$ while fixing the parameters of the other network. In our implementation, we employ the one-to-one ratio (1 discriminator iteration per generator update) which we empirically found can maintain the balance between the generator and discriminator. 
\begin{figure}[ht]
    \centering
    \includegraphics[width=\linewidth]{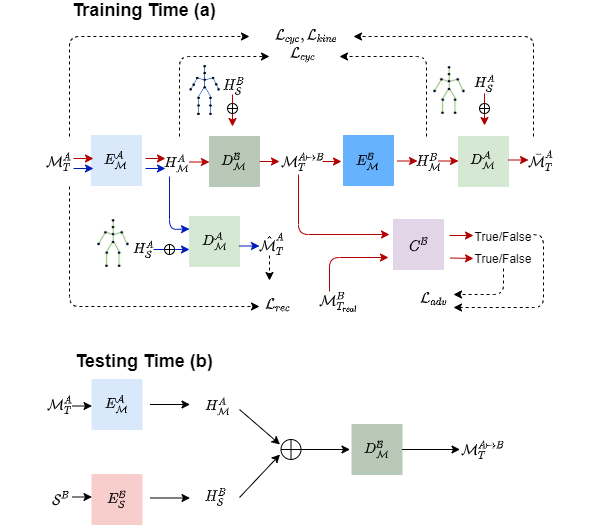}
    \caption{(a) Training process. The blue route shows the reconstruction process, which is used to train the encoder-decoder pairs. The red route shows the retargeting and cyclic retargeting process. (b) Testing Process. We discard the discriminator when inference.}
    \label{fig:training_process}
\end{figure}

Figure~\ref{fig:training_process} (a) shows the training process (we take the retargeting $A\rightarrow B$ as an example, $B\rightarrow A$ is symmetrically equivalent). We use the motion encoder $E_\mathcal{M}^\mathcal{A}$ to encode the source motion $\mathcal{M}_T^A$ of skeleton $A$ and then there are two branches (blue and red route in Figure~\ref{fig:training_process}). One is to reconstruct the motion $\hat{\mathcal{M}}_T^A$ by motion decoder $D_\mathcal{M}^A$ for training the encoder-decoder pair (blue route in Figure~\ref{fig:training_process}). The other branch (red route) is to utilize the decoder $D_\mathcal{M}^B$ with target skeleton code $H_\mathcal{S}^B$ to synthesize the retargeted motion $\mathcal{M}_T^{A\mapsto B}$. The motion discriminator $C^\mathcal{B}$ will be used to judge whether the retargeted motion is real or not. Since the latent motion code $H_\mathcal{M}^A$ is constructed by common body parts, we want it could be shared between different skeletal structure. Therefore, we extract the motion code $H_\mathcal{M}^B$ of the retargeted motion $\mathcal{M}_T^{A\mapsto B}$ by the corresponding motion encoder $E_\mathcal{M}^\mathcal{B}$, i.e, we cyclic retarget the motion $\mathcal{M}_T^{A\mapsto B}$ to $\bar{\mathcal{M}}_\mathcal{M}^A$. According to the idea, the $H_\mathcal{M}^A$ should be similar to $H_\mathcal{M}^B$. In addition to the latent code consistency, we also require the consistency in the original motion space when cycle retargeting, i.e., the motion $\bar{\mathcal{M}}_\mathcal{M}^A$ should be similar to the source motion $\mathcal{M}_T^A$. 

In the testing process (see Figure~\ref{fig:training_process} (b) ), the motion discriminator will be discarded, and we use source motion encoder $E_\mathcal{M}^\mathcal{A}$ and target skeleton encoder $E_\mathcal{S}^\mathcal{B}$ to synthesis the retargeted motion $\mathcal{M}_T^{A\mapsto B}$ through the motion decoder $D_\mathcal{M}^\mathcal{B}$. It is worth noting that the source and target structure can either be the same or different during training and testing.

\subsection{Loss functions}
According to the description of the training process, we are able to conclude the loss functions of our networks, \red{which are similar to SAN~\cite{aberman2020skeleton}.} We take the retargeting $A\rightarrow B$ as an example \red{to show the following loss terms:}

\noindent \textbf{Reconstruction loss.} In order to construct encoder-decoder pairs for different structure. We enforce the network to reconstruct the input samples. The loss term can be formulated as follow:
\begin{equation}
    \mathcal{L}_{rec} = \lVert \mathcal{M}_T^A - \hat{\mathcal{M}}_T^A \rVert ^2
    \label{eq:rec_loss}
\end{equation}
Where the $\hat{\mathcal{M}}_T^A$ is produced by its own motion encoder $E_\mathcal{M}^\mathcal{A}$, skeleton encoder $E_\mathcal{S}^\mathcal{A}$, and motion decoder $D_\mathcal{M}^\mathcal{A}$ (see the blue route in Figure~\ref{fig:training_process}(a) ). 

\noindent \textbf{Cycle consistency loss.} We regard each body part as our retargeting unit and suppose the motion encoder can produce the shared motion code $H_\mathcal{M}^A$. Therefore, to ensure the motion code we learned as the common motion features, we encourage the motion codes $H_\mathcal{M}^A$ and $H_\mathcal{M}^B$ to be as close as possible to each other in the latent space. At the same time, we encourage the motion $\mathcal{M}_T^A$ is similar to the cyclic retargeting motion $\bar{\mathcal{M}}_T^A$ in the original representation space by the following loss term:
\begin{equation}
     \mathcal{L}_{cyc} = \lVert H_\mathcal{M}^A - H_\mathcal{M}^B \rVert ^2 + \lVert \mathcal{M}_T^A - \bar{\mathcal{M}}_T^A \rVert ^2
    \label{eq:cycle_loss_latent}
\end{equation}

\noindent \textbf{Adversarial loss} Since we train our networks in an unsupervised manner, we need a discriminator to ensure the retargeted motion $\mathcal{M}_T^{A\mapsto B}$ looks plausible and falls onto the motion manifold of the corresponding structure $\mathcal{B}$. Therefore, we use the adversarial loss term described as follows:
\begin{equation}
    \mathcal{L}_{adv} = \lVert C^\mathcal{B}(\mathcal{M}_{T}^{B_{real}})\rVert ^2 + \lVert 1-C^\mathcal{B}(\mathcal{M}_T^{A\mapsto B})\rVert ^2
    \label{eq:adv_loss}
\end{equation}
Where the $\mathcal{M}_{T}^{B_{real}}$ are real examples of structure $\mathcal{B}$. When training the discriminator alone, we detach the retargeted motion generated by the motion generator from the calculation graph and maximize the loss function~\ref{eq:adv_loss} to encourage the discriminator to distinguish between real data and synthesized motions. While training the generator, we minimize the  function~\ref{eq:adv_loss} as well as other loss terms to fool the discriminator for obtaining more realistic retargeting results.

\noindent \textbf{Kinematic loss.} The motion representation $\mathcal{M}_T^A$ is sufficient to determine a motion when combined with offsets $S^A$. But, sometimes people are more interested in the correspondence in Cartesian space, especially the position of end-effectors such as foot contacts. Therefore, we transform the local rotation of each joint into 3D coordinates in Cartesian space by forward kinematics (FK). Then we encourage the networks to minimize the joint position errors when reconstruction and cyclic retargeting.
\begin{equation}
    \mathcal{L}_{kine} = \lVert FK(\mathcal{M}_T^A) - FK(\bar{\mathcal{M}}_T^A)\rVert ^2 + \lVert FK(\mathcal{M}_T^A) - FK(\hat{\mathcal{M}}_T^A)\rVert ^2
\end{equation}
where the $\bar{\mathcal{M}}_T^A$ and $\hat{\mathcal{M}}_T^A$ are cyclic retargeted and reconstruction motions, respectively.

The final loss function can be summarized as follow:
\begin{equation}
    \mathcal{L}_{total} =  \lambda_{1}\mathcal{L}_{rec} + \lambda_{2}\mathcal{L}_{cyc} + \lambda_{3}\mathcal{L}_{kine} + \lambda_{4}\mathcal{L}_{adv}
    \label{eq:total_loss}
\end{equation}
In our implementation each $\lambda$ value is set to 1, 2.5, \redd{$10^{2}$}, 1, respectively.

\subsection{Implementation Details}
Our architecture is trained for 1000 epochs with a batch size of 128 under the PyTorch platform. We use Adam optimizer~\cite{2014Adam} with $\beta_1=0.9$ and $\beta_2 = 0.999$ to train the generator and the discriminator, whose learning rates are both set to $10^{-3}$. The training time is about 12 hours with NVIDIA RTX 3090ti.

We localize motion frames by rotating the root on the y-axis so the character is always facing one direction(z-axis positive in our case) which is proven to be beneficial for the convergence of neural networks~\cite{holden2017phase}.
% For humanoid characters, the facing direction of the character is computed by averaging the vector between the hip joints and the vector between the shoulder joints. For quadrupeds in dataset~\cite{zhang2018mode}, the facing direction is determined by the root-spine bone projection onto the ground. 
After processing, the information of global translation, as well as the y rotation of the root, is contained in the velocity vector $\bar{V}$. To enable training the network in batches, we cut the motions in the datasets into motion clips with fixed length $T=64$. All examples of animation are downsampled at a rate of 30 fps, so the clip length $T$ is able to make the networks distinguish most types of motion without affecting the training efficiency. We normalized each motion data $\mathcal{M}_T$ by z-score as follow:
\begin{equation}
    \mathcal{M}_T = \frac{(\mathcal{M}_T - \mu_\mathcal{M})}{\sigma_\mathcal{M}}
\end{equation}
Where the $\mu_\mathcal{M}$ and $\sigma_\mathcal{M}$ represent the mean and standard deviation of all motion data $\mathcal{M}_T$ in the training dataset, respectively. 

\section{Retargeting between Humanoid Skeletons}\label{sec:experiments}
\begin{figure*}[ht]
    \centering
    \includegraphics[width=\linewidth]{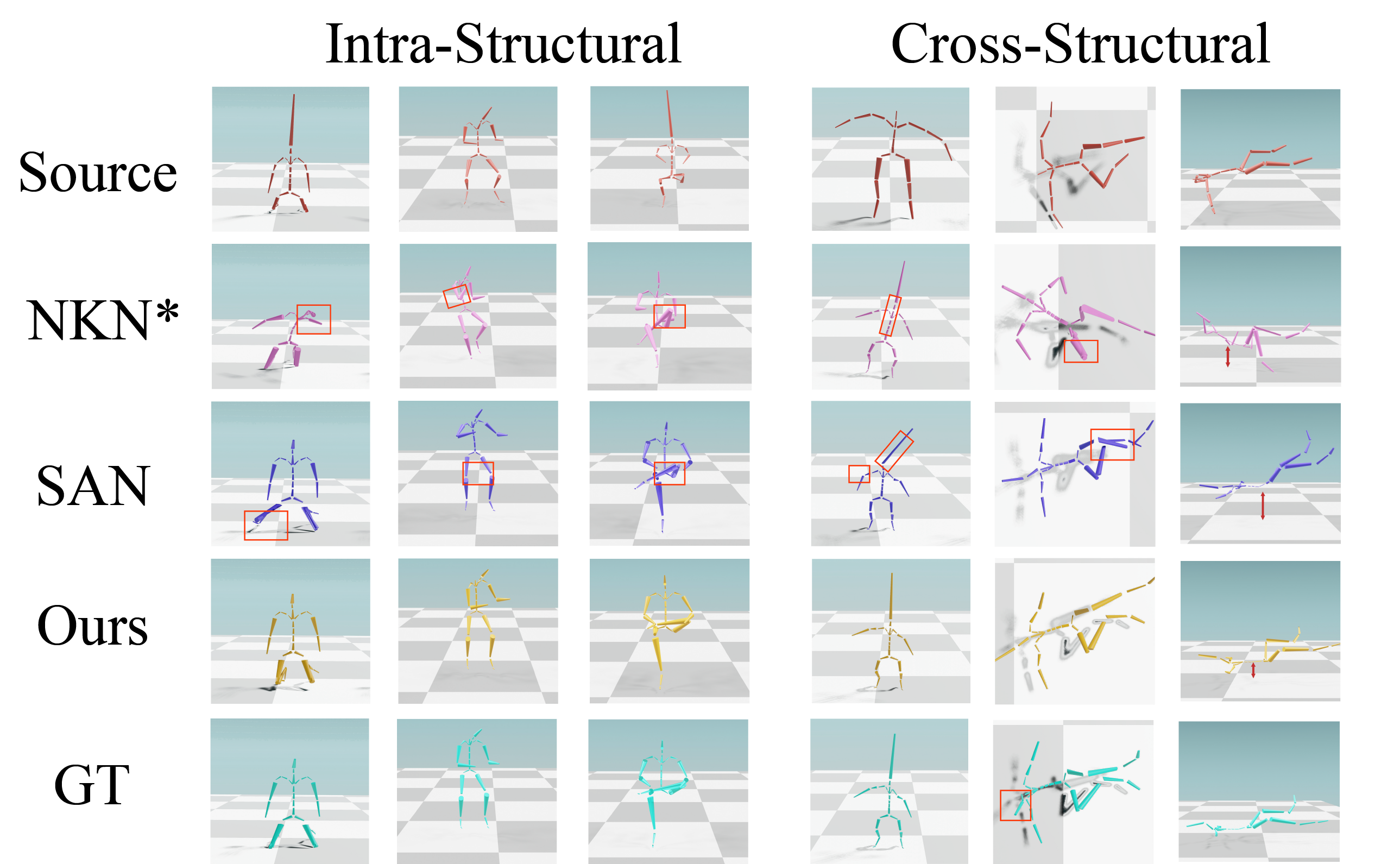}
    \caption{\red{Qualitative comparisons between our method, SAN, modified NKN, and the corresponding ground truth. The first row displays the source poses while the other rows show the results of the Retargeting. Flaws in the results are marked by red rectangular boxes and arrows.}}
    \label{fig:intra_cross_structural}
\end{figure*}
In this section, 
% we carry out the experiments compared with other retargeting methods on Mixamo~\cite{mixamo} dataset, 
we mainly evaluate the effectiveness of the proposed method in motion retargeting between humanoid skeletons. For more qualitative results, please refer to the supplementary video.

\subsection{Problem Setting}
Humanoid retargeting is divided into two cases based on the differences between the source and target structure. One is intra-structural retargeting, i.e., the target skeleton has the same structure and joint number, but different bone proportions. The other is cross-structural retargeting, where the target skeleton has a similar topology to the source skeleton but a different number of joints. To take both cases into account, we follow the work~\cite{aberman2020skeleton} and divide the characters into two groups, $\mathcal{A}$ and $\mathcal{B}$, each containing skeletons with the same structure but different bone proportions. Compared to structure $\mathcal{B}$, the skeletons of structure $\mathcal{A}$ contain extra joints on each limb(left/right arms, left/right leg, spine, and neck).

\noindent \textbf{Dataset.} 
The Mixamo~\cite{mixamo} is a 3D motion dataset with rich motion types and contains more than 2000 sequences performed by 29 distinct humanoid characters, of which structure $\mathcal{A}$ includes 24 characters and structure $\mathcal{B}$ contains 5 characters. In our experiments, we select 20 characters in the group $\mathcal{A}$ and 4 characters in the group $\mathcal{B}$ for training, and the remaining characters are used for testing. In addition, we follow the literature~\cite{villegas2018neural, aberman2020skeleton} to clip the fingers of each character and keep the main limb joints for simplification. 

\noindent \textbf{Comparison Methods.}
The methods we compared are NKN~\cite{villegas2018neural}, PMnet~\cite{lim2019pmnet} and SAN~\cite{aberman2020skeleton}. NKN is a pioneering work in deep learning-based motion retargeting. It uses Recurrent Neural Network(RNN) to model the temporal relationship of motion and combine the skeleton offsets for intra-structural retargeting. PMnet~\cite{lim2019pmnet} learns frame-by-frame poses and overall movement separately for improving the retargeting accuracy. The most similar work to ours is SAN~\cite{aberman2020skeleton}, it extracts common skeletal codes by skeleton-based graph convolution which can be applied to skeletons with different structure. We will demonstrate in the following sections that our attention-based spatial modeling at the body part level is more conductive and flexible to the task of motion retargeting. 

\subsection{Experiments and Evaluation}~\label{sec:mixamo_experiment}
% The basic retargeting requires adapting motion between skeletons with same structure but different proportions. Thus, 
We use skeletal structure $\mathcal{A}$ to evaluate the intra-structural retargeting. At this time, our modules in the architecture degenerate to the special case, that is, the structure of the target skeleton is the same as that of the source skeleton, both of which are $\mathcal{A}$. The different skeletons $[A_1, A_2, ..., A_n]\in \mathcal{A}$ are distinguished by offsets $\mathcal{S}^{A_i}$ which represents the skeleton topology and bone proportion. 
\begin{figure*}[ht]
    \centering
    \includegraphics[width=\linewidth]{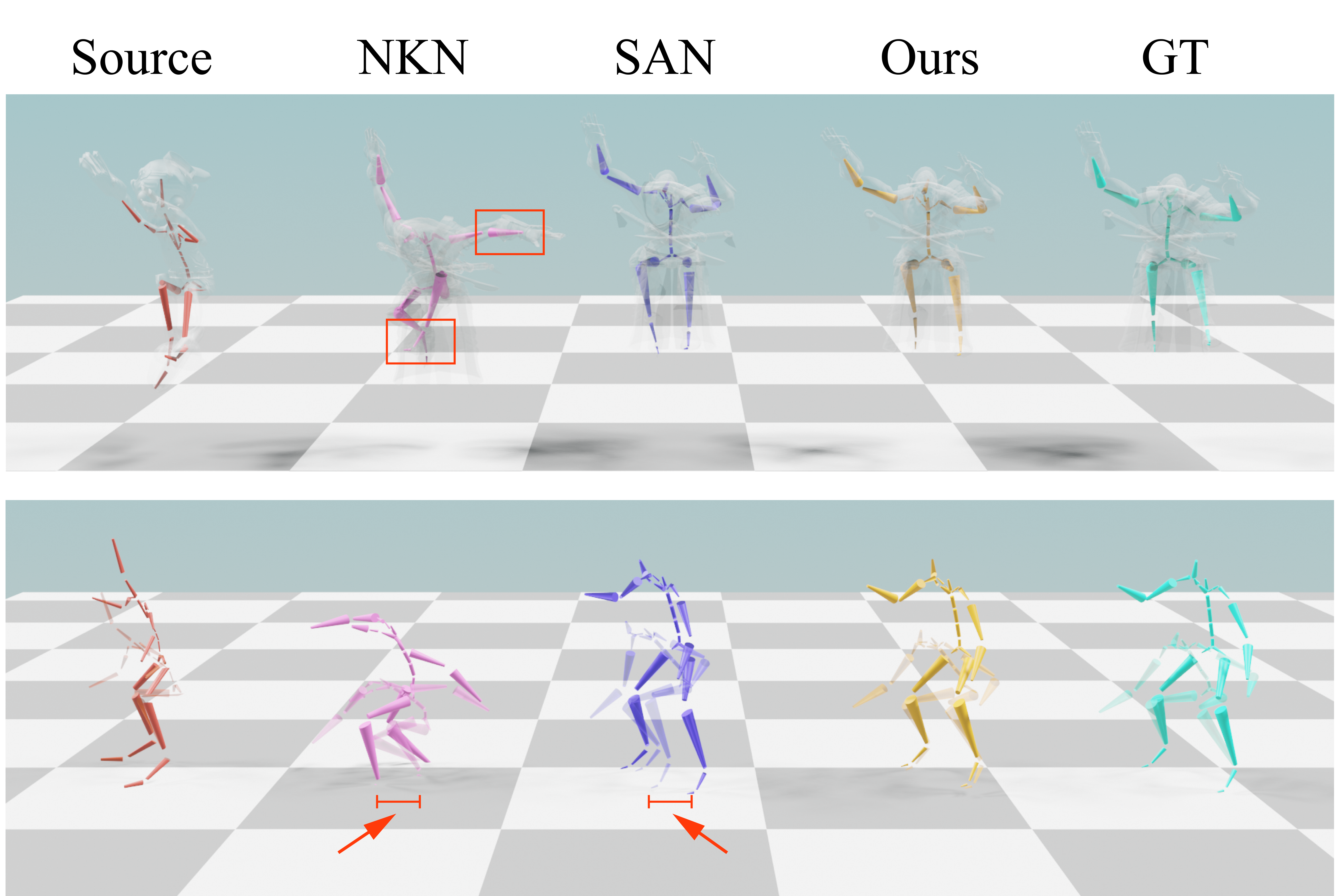}
    \caption{A jumping-to-land motion clip. We show the retargeting results of NKN, SAN, and our method along with the corresponding ground truth. In the first row, we overlaid the transparent mesh on the skeleton. In the last row, we show the pose when standing(solid) and the pose when just touching the ground (transparent). Retargeting flaws or server foot sliding are marked with red rectangles and arrows.}
    \label{fig:foot_sliding}
\end{figure*}

To evaluate the cross-structural retargeting, we allow the motions to be adapted between structure $\mathcal{A}$ and $\mathcal{B}$. Since the target skeleton has a different number of joints than the source skeleton, both the vanilla NKN and PMnet models are no longer applicable to this case due to the requirement of the same dimension between the input and output skeletons. Therefore, we modified these models by retraining the encoder-decoder pair of each structure and forcing the latent code dimension to be the same in order to share among different structure. The modified models are denoted by NKN$^*$ and PMnet$^*$, respectively. 

\red{Columns 1-3 in Figure~\ref{fig:intra_cross_structural} show the comparison examples} of intra-structural retargeting while the next \red{three columns present} the cross-structural retargeting results. From the comparison, we can find our method achieves more stable and accurate results. In particular, because of the unsupervised training manner, we are able to produce reasonable results when there are little flaws in the ground truth (see the \red{middle column of the cross-structural retargeting in Figure~\ref{fig:intra_cross_structural}})
\begin{table}[ht]
    \centering
     \caption{Quantitative Evaluation on Mixamo Dataset. 
     % We Compare Our Approach with NKN, PMnet and SAN. The Modified NKN$^*$ and PMnet$^*$ are Tested for Cross-structural Scenario (detailed in sec~\ref{sec:mixamo_experiment} ). 
     We Report the Mean Per Joint Position Error(MPJPE) over Test Clips, Normalized by the Skeleton's Height (multiplied by 10$^3$, similar with~\cite{aberman2020skeleton}).}
    \begin{tabular}{c||c||c}
        \hline
         ~ & Intra-Structural  &  Cross-Structural \\
         \hline\hline
        Copy rotations & 8.86 & N/A\\
         
        NKN/NKN$^*$ & 5.84 & 7.36\\

        PMnet/PMnet$^*$ & 4.93 & 6.88\\

        SAN & 2.76 & 2.25\\
    
        Ours& \textbf{0.50} & \textbf{1.62}\\
        \hline
                
    \end{tabular}
    \label{tab:mse_maximo}
\end{table}

\subsubsection{Quantitative Evaluation}
For quantitative evaluation, we use mean per joint position error(MPJPE) as a metric and compare the generated results with ground truth over $I$=106 test motion clips. The MPJPE formula is described as follows: 
% The quantitative results of humanoid retargeting are reported 
\begin{equation}
    \mathcal{E} = \frac{1}{I\vert \mathcal{C}_s \vert \vert \mathcal{C}_t \vert h_t}\sum_{i=1}^{I}\sum_{a\in \mathcal{C}_s}\sum_{b\in \mathcal{C}_t} \lVert FK(\mathcal{M}^{s\mapsto t}_{T_i}) - FK(\mathcal{M}^t_{T_i})\rVert^2
\end{equation}
Where the $\mathcal{M}_{T_i}^{s\mapsto t}$ and $\mathcal{M}_{T_i}^{t}$ are the retargeted motion from skeleton $s$ to $t$ and ground truth motion from the Mixamo dataset, respectively. $FK$ is a forward kinematic function for transferring the joint rotations to global positions. $i$ denotes the index of the test motion examples. $\mathcal{C}_s$ and $\mathcal{C}_t$ represent the source and target skeleton set. In the intra-structural retargeting case, the set $\mathcal{C}_t$ is the same as $\mathcal{C}_s$, i.e., both are the 4 testing skeletons in structure $\mathcal{A}$. In cross-structural retargeting, the source set $C_s$ contains only 1 test skeleton in the group $\mathcal{B}$, and $C_t$ contains the 4 skeletons in the group $\mathcal{A}$. In order to eliminate the error magnitude caused by different skeleton sizes, we divide the joint position error by target skeleton height $h_t$. 

The quantitative results are reported in Table~\ref{tab:mse_maximo}. Since the target skeleton has one-to-one joint correspondence with the source skeleton in intra-structural retargeting, we additionally compute the result of Copy rotations as a baseline, i.e., directly copying each joint rotation to the target skeleton. From the table, we can see the performance of NKN and PMnet is far behind SAN and our method since they feed the networks with a simple concatenation of whole-body joint features, without modeling the spatial characteristics of the skeletons. Our approach outperforms all competing methods in both intra-structural and cross-structural retargeting, attributing to the expressive power of the attention mechanism and the body part strategy, which will be further discussed.
\begin{figure}[ht]
    \centering
    \includegraphics[width=\linewidth]{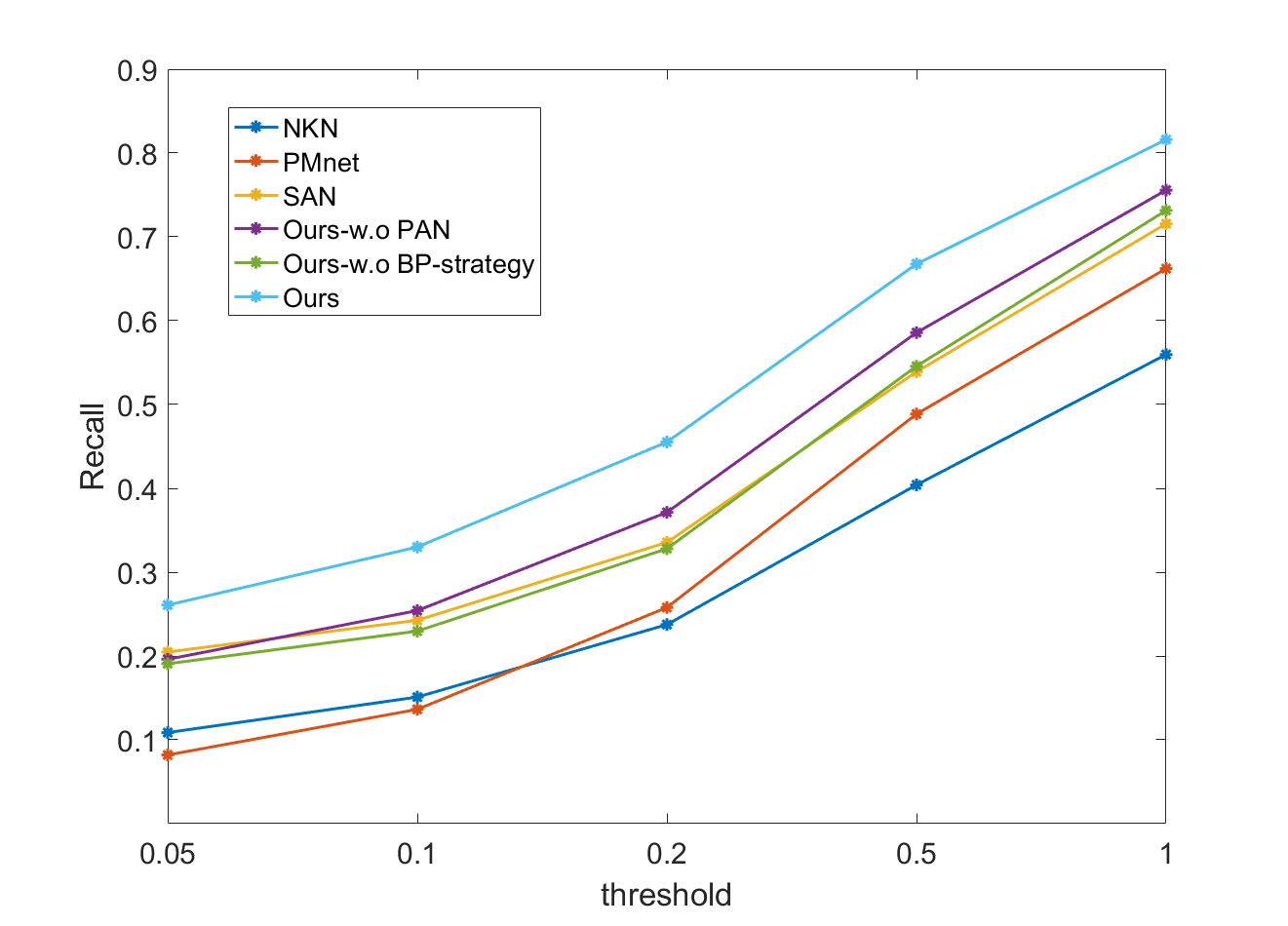}
    \caption{The foot-contact recall of different methods. Given different foot velocity thresholds, we compute the foot-contact recall by equation~\ref{eq:contact_recall}. A higher value indicates the foot movement is closer to the ground truth.}
    \label{fig:contact_recall}
\end{figure}

\begin{figure*}[ht]
    \centering
    \includegraphics[width=\linewidth]{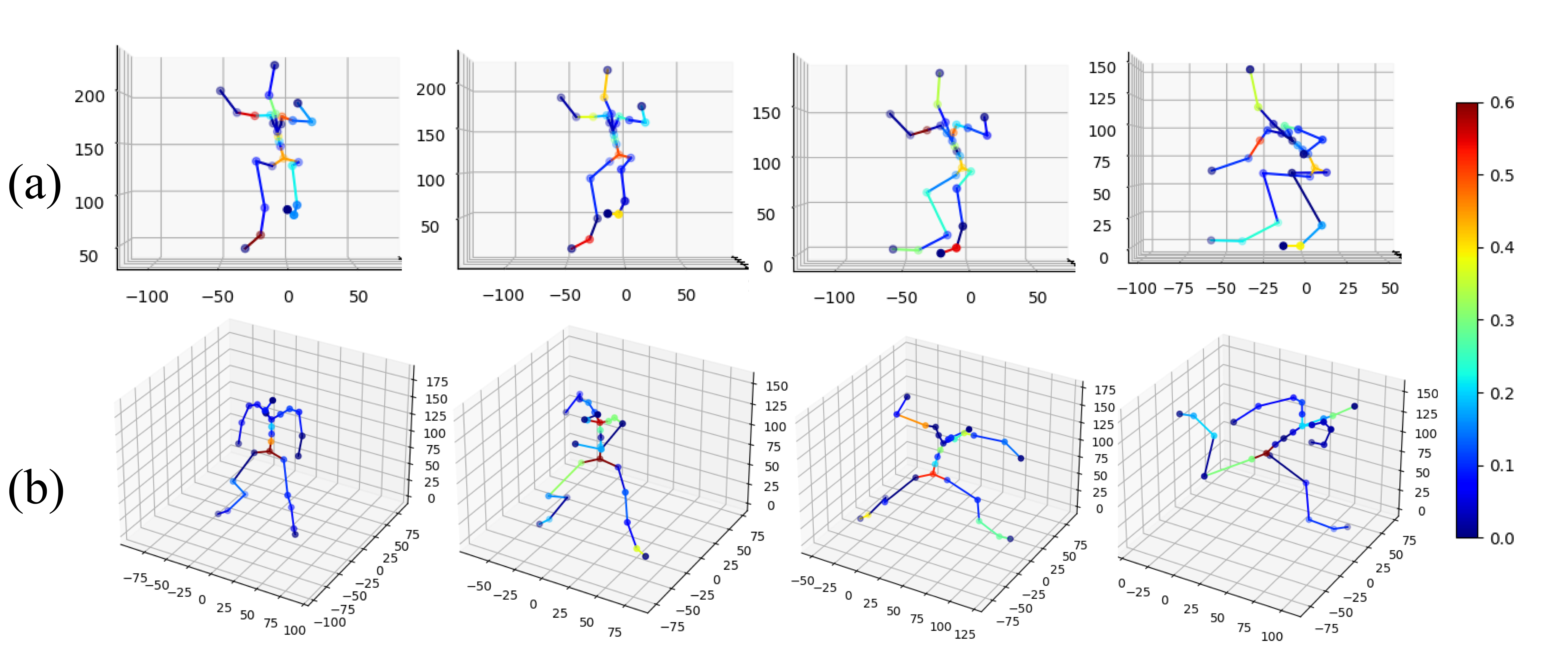}
    \caption{Visualization of the pose-aware attention. We visualize two motion sequences (a) and (b), temporally ordered from left to right. The contribution of each joint to the corresponding body part is described by a heat map. Notably, we overlaid the attention weights of root rotation and root velocity.}
    \label{fig:attention_map}
\end{figure*}
\subsubsection{Evaluation of Foot-contact Recall}
In addition to the overall accuracy of motion retargeting, the movement of the end-effectors usually plays an important role, especially when considering contacts with the ground, which is crucial to our perception of human motions. However, the results produced by neural network systems often do not satisfy the contact constraints. Therefore, there is some literature like~\cite{villegas2021contact} utilizes numerical optimization in the hidden space to iteratively improve the solution. We believe that a good neural network prediction provides a better initial solution for the optimization process and can speed up the convergence of the algorithm, resulting in more visually plausible motions. We conduct experiments and visualizations to demonstrate that our pose-aware attention mechanism can provide more stable results, mitigating foot-sliding artifacts. 

Figure~\ref{fig:foot_sliding} shows a jumping-to-land motion clip. We can see the source motion as well as the ground truth without any foot sliding after landing on the ground, but there are artifacts in each of the predicted results. The comparison in the last row of Figure~\ref{fig:foot_sliding} reveals that our method achieves more stable results with slight sliding. Although we do not specifically consider contact information during the training, the proposed method adaptively focuses attention on the foot joint within the leg body part due to the pose-aware attention mechanism, which is well visualized in Figure~\ref{fig:attention_map}(a).

The visualization shows the contribution of each joint to the corresponding body part in different postures. We take the output of $softmax$ in equation~\ref{eq:attention_function} at the first attention layer and draw the attention weights of the joint embeddings $X_t\in \mathbb{R}^{(J+1)\times d}$ to body part tokens $Y\in \mathbb{R}^{N\times d}$ by heatmap. The color near the top of the color bar indicates that the joint contributes more to its corresponding body part and vice versa. It is noticed that \red{the feature of root velocity will participate in the calculation of attention in each body part since the root velocity is important for distinguishing the motion types. We will take the maximum weight value when visualizing if a joint belongs to more than one body part and} the color of the root is overlaid with the root velocity "joint".

For different motion poses, the attention weights computed by the network vary considerably. For example, in sequence (b) of Figure~\ref{fig:attention_map}, we can observe that in the case of stationary pose (the first frame), \red{the PAN focuses more on the root/root velocity and the weight contributed by each joint to the corresponding body part do not differ much since the pose remains stationary.} This meets our intuition that \red{the root velocity is important since it can distinguish multiple motion categories and each joint contributes similar weight at rest pose.} However, it is not a static rule, attention maps for other poses in Figure~\ref{fig:attention_map} (b) show that our pose-aware attention mechanism has the ability to dynamically extract the spatial features between joints.

To further demonstrate our attention mechanism can mitigate foot-sliding artifacts, we use the foot-contact recall as a metric and compare our approach with other methods. Firstly, we collect all of the contact frames for each foot in ground truth motion by velocity threshold, described as follows:
\begin{equation}    
    c_{gt}(t;\epsilon) = 
    \begin{cases}
    1 &\mbox{if $\lVert P_{t,j}^{gt} -  P_{t-1,j}^{gt}\rVert^2 < \epsilon$}\\
    0 & \mbox{otherwise}
    \end{cases}
\end{equation}
where the $P_{t,j}^{gt}$ represents the ground-truth position of foot joint $j$ at time $t$. We collect all the contact frames to form a set $\mathcal{C}_{gt} = [t_1, t_2, ..., t_n]$ and then check the retargeted motion for foot contacts on these frames by the following equation:
\begin{equation}
    c_{tar}(t_i;\epsilon) = 
    \begin{cases}
    1 &\mbox{if $\lVert P_{t_i,j}^{tar} -  P_{t_i-1,j}^{tar}\rVert^2 < \epsilon$}\\
    0 & \mbox{otherwise}
    \end{cases}
    \quad t_i \in \mathcal{C}_{gt}
\end{equation}
Finally, we are able to calculate the foot-contact recall by:
\begin{equation}
    \mathcal{R}_{\epsilon} = \frac{\sum_{t_i\in \mathcal{C}_{gt}}c_{tar}(t_i;\epsilon)}{\sum_{t\in T}c_{gt}(t;\epsilon)}
    \label{eq:contact_recall}
\end{equation}
We plot the line graph (see Figure~\ref{fig:contact_recall}) given different threshold $\epsilon$ and can find that our full approach outperforms all competing methods at each threshold, further demonstrating the proposed method can produce more stable results. We also conduct an ablation study on this metric, which will be discussed in section~\ref{sec:ablation}.

\subsubsection{Motion Retargeting from Human3.6M}\label{sec:h36m}
\red{To demonstrate the generalization of our method, we present motion retargeting from the Human3.6~\cite{ionescu2013human3} characters to Mixamo's skeletons using the model trained from Mixamo data only, which means that our architecture can retarget unseen motions performed by unseen skeletons to unseen/seen skeletons. We use the ground truth 3D motions (joint rotations for our method and Copy rotation, joint positions for NKN) from the human3.6M dataset, and downsample them to 30fps. }

\red{In order to enable the encoder trained on the Mixamo structural $\mathcal{B}$ (22 joints) to receive the articulated motion from Human3.6M (17 joints), we did joint mapping similar to NKN, which duplicates the joint positions in Human3.6M to corresponding Mixamo joints. Since our method receives joint rotations as input, we map the Spine rotation (H36M) into the Spine(Mixamo) and set the rotation of Spine1 (Mixamo) to zero. At the same time, the offset of Spine1 will also be set to zero, thus creating a zero-length bone in the test time. A similar operation will be applied to follow mapping pairs: LeftShoulder into LeftShoulder and LeftArm, RightShoulder into RightShoulder and RightArm, LeftFoot into LeftFoot and LeftToeBase, RightFoot into RightFoot and RightToeBase. Thus, our motion encoder trained for structural $\mathcal{B}$ can receive the motions of human3.6M dataset.} 

\begin{figure}
    \centering
    \includegraphics[width=\linewidth]{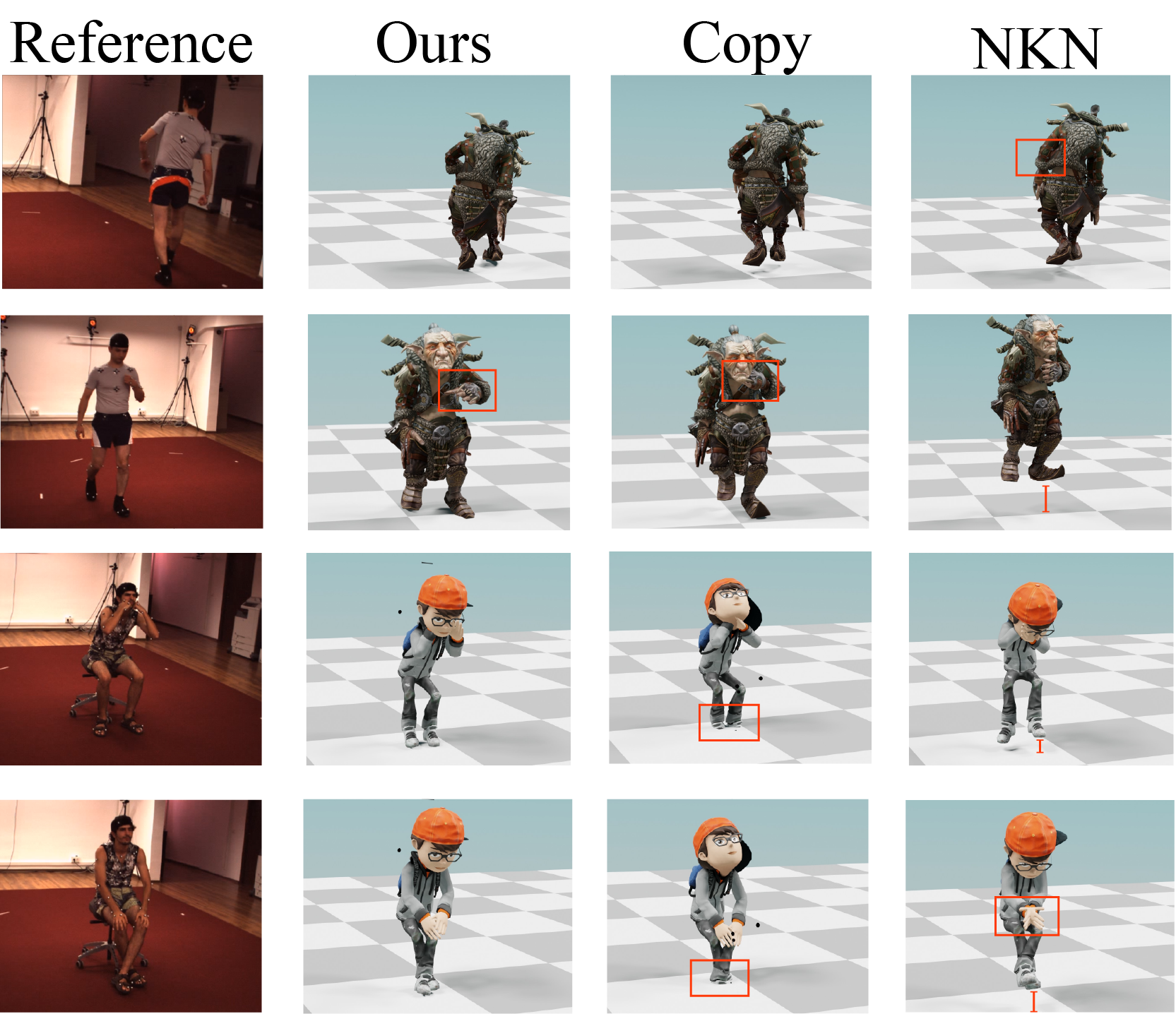}
    \caption{Qualitative results of retargeting from Human3.6M motions to Mixamo skeletons (the Mixamo skeletons in rows 1-2 are unseen during training, while the ones in rows 3-4 are seen). We use ground truth 3D motions of the Human3.6M dataset as the source motions.}
    \label{fig:human36m}
\end{figure}

\red{As shown in Figure~\ref{fig:human36m}, our method can generalize to unseen articulated motions and outperform NKN and Copy Rotation in terms of realism and plausibility. For example, we can see that directly copying the joint rotations of the human3.6m' motion to a target skeleton may produce unreasonable poses due to the structural differences between the source and target skeletons.}

\red{For NKN, since the method retargets motions in an autoregressive manner, the errors will accumulate as the motion sequence becomes longer, resulting in the generated character motion floating in the air (Row 2-4 in Figure~\ref{fig:human36m}). \redd{In contrast, our temporal modeling is based on 1D convolution, which makes the model more stable in long-term retargeting. Please refer to the supplementary video for more qualitative results.}}

\redd{To quantitatively evaluate each retargeting method, we randomly select the source motion clips (300 frames per clip) from the Human3.6M datasets and retarget them to the unseen characters of Mixamo, then allow users to score them for evaluation. Specifically, We run our user study on a total of 20 users and choose Mousey, Mremireh, and Vampire as test characters (neither our model nor NKN saw them during training). We assign 5 retargeting clips to each character, making a total of 15 questions. For each question, we randomly place the retargeting results produced by each method on the page, labeled A, B, and C. The participants are asked to grade the similarity between the retargeted and source motions on a scale of integers from 1 to 5, with ”1” denoting completely dissimilar and ”5” is completely similar. Table~\ref{tab:h36m_user} shows the results of the user study. We can find that our method outperforms the NKN and Copy Rotation in terms of the scores and has a smaller standard deviation, which means that our method can achieve more accurate and robust retargeting. }
\begin{table}[ht]
    \caption{User Study of Retargeting from Human3.6M to Characters of Mixamo.}
    
    \centering
    \begin{tabular}{c|c|c}
        \hline
         Copy Rotation & NKN & Ours  \\
         \hline
           3.06$\pm$ 1.07 & 3.69$\pm$ 1.04 & \textbf{3.78}$\pm$ \textbf{0.98}  \\  
         \hline 
                
    \end{tabular}
    \label{tab:h36m_user}
\end{table}

\section{Retargeting between biped and quadruped}\label{sec:quadruped}
\begin{figure}[ht]
    \centering
    \includegraphics[width=\linewidth]{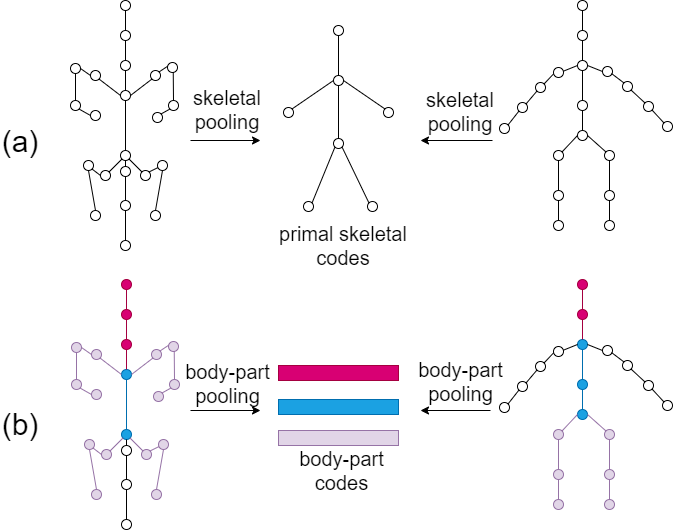}
    \caption{Different corresponding strategies between our method and SAN. (a) SAN extracts the primal skeleton features based on the neighborhood relationship of the joints. (b) we construct the shared latent space of the source and target skeletons in terms of body parts. }
    \label{fig:san_vs_ours}
\end{figure}

Our retargeting framework is able to define more flexible correspondence due to the body part retargeting strategy \red{and pose-aware attention mechanism}. To demonstrate this property, we design motion retargeting between quadrupeds and bipeds (see Figure~\ref{fig:san_vs_ours} left and right sides) on the more challenging datasets lafan1~\cite{harvey2020robust} and quadrupeds~\cite{zhang2018mode}. 
\subsection{Problem Setting}
\noindent \textbf{Datasets.} 
The lafan1~\cite{harvey2020robust} is a human motion dataset that includes walking, running, sitting, sprinting, fighting, etc. There are 77 unique motion sequences, and we choose the locomotion clips \red{(i.e., BVH files starting with the keywords aiming, run, walk, and sprint)} because the motion mode of quadrupeds is relatively simple (mainly locomotion). \red{We use subject 1 for testing and the rest for training.} The quadruped dataset~\cite{zhang2018mode} consists of 52 unique dog motion sequences including idle, walk, run, sit, stand, and a few jumps. We downsample the frame rate to 30 fps, consistent with the lafan1 dataset. \red{For train-test split, we split the whole 52 motion files randomly and make sure the train-test ratio is close to the bipedal dataset split. Please refer to Appendix B for detailed information.}

Figure~\ref{fig:san_vs_ours} shows the differences between our method and SAN~\cite{aberman2020skeleton} in terms of the corresponding strategies for bipedal and quadrupedal skeletons. The two skeletons have different structure, but the main limbs are topologically similar. The core idea of SAN is to extract primal skeletal codes through multiple pooling operations (see Figure~\ref{fig:san_vs_ours} (a)). However, this strategy leads to a correspondence between the arm motions of bipeds and the foreleg motions of quadrupeds, which is semantically implausible. \HL{Instead, we define $N$=3 common body parts of bipeds and quadrupeds (see Figure~\ref{fig:san_vs_ours} (b)) rather than a primal skeleton}. Since it is difficult to find a spatial correspondence between the leg part of this two structure, we directly pack the biped's two legs to construct a correspondence with the four legs of the quadruped to ensure correctness on the semantic level. It is worth noting that we do not encode the motion of the arms of biped or the tail of quadruped, because these body parts are difficult to construct a correspondence between the structure. \red{It means that the decoders corresponding to each structure receive only the codes of common body parts and decode the whole-body motions with these partial part features. }

\red{During the training process, \redd{our PAN calculates the attention weights for each body part, which function similarly to the gating network in MANN~\cite{zhang2018mode} to learn to distinguish between different actions and motion states by clustering. Meanwhile, since we use temporal convolutional layers to compress the motion, our architecture has a wider perceptual field than frame-by-frame methods and thus does not take action labels as input.} The naturalness of the generated whole-body motion will be judged by the motion discriminator, and the $\mathcal{L}_{adv}$ is used to force the motion decoder to generate natural and reasonable motions. Meanwhile, we will constrain only the common body parts in $\mathcal{L}_{rec}$, $\mathcal{L}_{cyc}$, $\mathcal{L}_{kine}$. }\redd{We additionally add velocity constraints to the biped-quadruped retargeting setting as follow:  
\begin{equation}
    \mathcal{L}_{vel} = \lVert \frac{V_s}{\lVert V_s\rVert}(\frac{\lVert V_s\rVert - v_{smin}}{v_{smax} - v_{smin}}) - \frac{V_t}{\lVert V_t\rVert}(\frac{\lVert V_t\rVert - v_{tmin}}{v_{tmax} - v_{tmin}}) \rVert ^2
    \label{eq:rec_loss}
\end{equation}
Where the $V_s$ and $V_t$ represent the root joint velocity of source and target skeletons. $v_*min$ and $v_*max$ indicate the minimum and maximum velocity scale in
the training datasets, respectively. 
This loss term forces the velocity to be mapped in proportion which can help us align the motion manifolds of the two morphologies and circumvent unreasonable retargeting. The coefficient of the velocity loss term is $\lambda_{vel} = 10^3$ and other coefficients are the same as in equation~\ref{eq:total_loss}. 
}
\begin{figure}[ht]
    \centering
    \includegraphics[width=\linewidth]{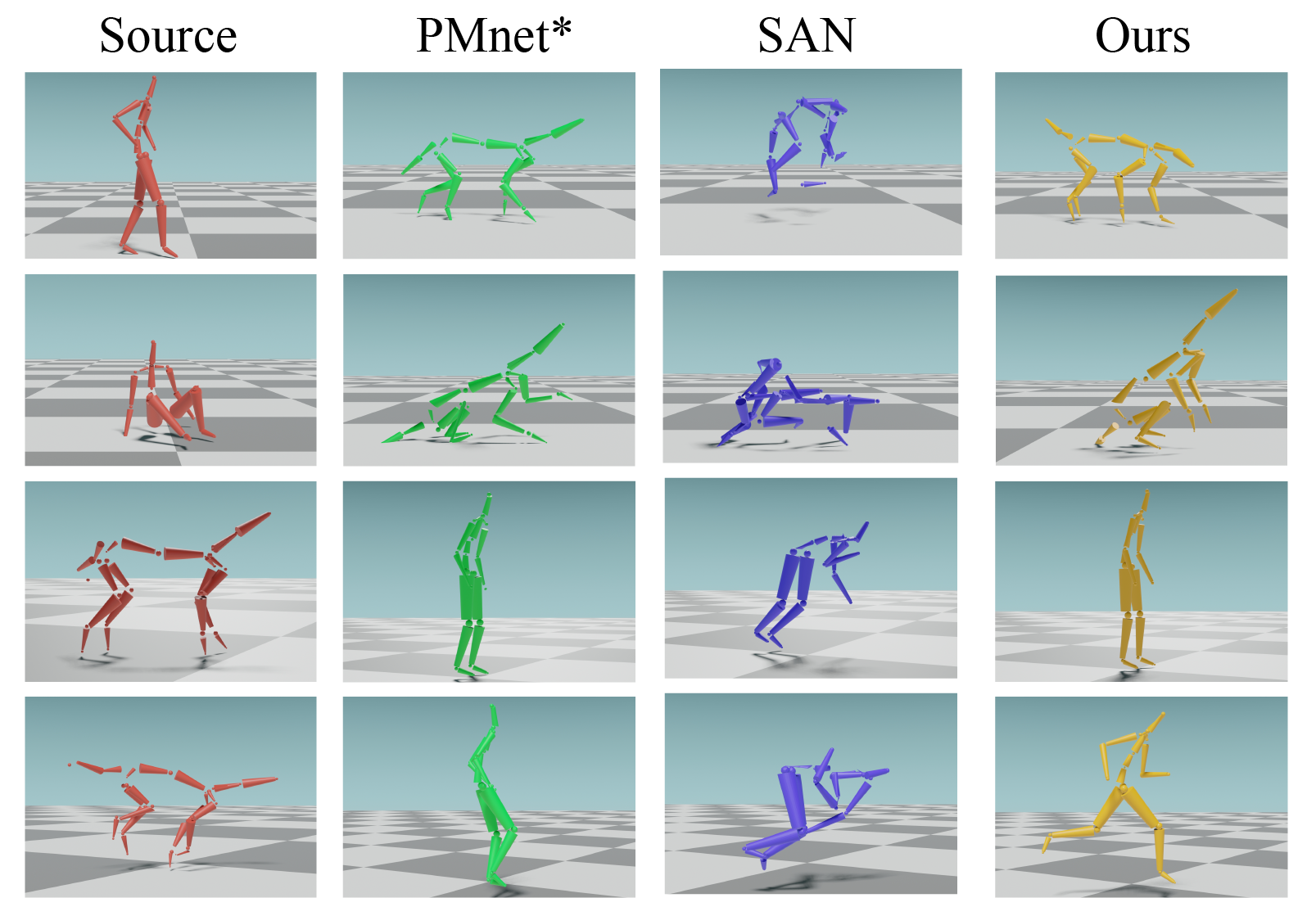}
    \caption{Qualitative results of retargeting between bipedal and quadrupedal. The leftmost column shows the source poses and the columns on the right represent the retargeting results of modified PMnet, SAN, and our method. 
    \red{The results are all evaluated on the test set.} }
    \label{fig:qualitative_quadruped}
\end{figure}

\subsection{Experiments and Evaluation}
When retargeting the quadruped motion to the bipedal character, the results produced by SAN all present a "bent" posture (see the last two rows of Figure~\ref{fig:qualitative_quadruped}) because of the unreasonable correspondence. Similarly, SAN fails on retargeting from biped to quadruped (see the first two rows of Figure~\ref{fig:qualitative_quadruped}) where only two legs are in contact with the ground and the poses are very unnatural. The qualitative results in Figure~\ref{fig:qualitative_quadruped} show our method can produce more realistic-looking results, demonstrating our body-part corresponding strategy is more reasonable. We also show the results of PMnet$^*$, which performs between our method and SAN since it does not consider any spatial relationship between these two skeletons but only models the motion along the temporal dimension.

\subsubsection{Quantitative Evaluation}
For quantitative evaluation, since the lafan1 and quadruped datasets have no motion pairs, we evaluate the retargeting by two metrics: Fréchet Inception Distance(FID) and user study. 
\begin{table}[ht]
    \caption{We Evaluate the Fréchet Inception Distance(FID) on Lafan1 and Quadruped Dataset.}
    \centering
    \begin{tabular}{c||c||c}
        \hline
         ~ & Quadruped$\rightarrow$Biped  & Biped$\rightarrow$Quadruped \\
         \hline\hline
        NKN$^*$ & 93.81 & 169.47\\
        PMnet$^*$ & 62.55 & 217.11 \\
        SAN & 376.36 & 763.89 \\
        Ours& \textbf{51.68 } & \textbf{132.76}\\
        \hline
                
    \end{tabular}
    \label{tab:fid}
\end{table}

\begin{table*}[ht]

    \caption{Mean Subjective Ratings with Standard Deviation. We Ask the Users to Grade the Similarity Between the Retargeted and Source Motions.}
    \centering
    \begin{tabular}{c|c|c|c||c|c|c}
        \hline
         \multirow{2}{*}{~}& \multicolumn{3}{c||}{Biped$\rightarrow$ Quadruped} & \multicolumn{3}{c}{Quadruped$\rightarrow$ Biped}\\
         \cline{2-7}
          & idle & move & sit & idle & move & sit \\
         \hline
         NKN$^*$ &1.94$\pm$0.75 &2.94 $\pm$0.56 &1.53$\pm$0.51 &2.88$\pm$0.99 &2.18$\pm$0.64 &2.06$\pm$1.25  \\
         PMnet$^*$ &2.35$\pm$0.70 &3.06 $\pm$ 0.75 &1.82$\pm$0.73 & \textbf{3.82 $\pm$ 0.88} &3.89$\pm$0.86 &2.65$\pm$1.37 \\ 
         SAN  &1.59$\pm$1.18 &1.41$\pm$1.00 &2.12$\pm$1.11 &1.29$\pm$0.47 &1.35$\pm$0.61 &2.24$\pm$1.20\\ 
         Ours  &\textbf{4.29$\pm$0.69} &\textbf{4.53 $\pm$ 0.62} &\textbf{4.53$\pm$0.51} &3.76$\pm$1.03 &\textbf{4.06$\pm$0.75} &\textbf{3.76$\pm$1.03} \\

    \end{tabular}
    \label{tab:user_study}
\end{table*}
FID is widely used to evaluate the overall quality of generated motion~\cite{2020Action2Motion}, which depicts how similar the generated motions are to the real motions by comparison of the deep feature distributions. As we know, similar motions should have similar hidden features in the hidden layer of the network, so we pre-train an auxiliary autoencoder~\cite{holden2015learning} to help us compute the FID scores (i.e. all the retargeted motions and real motions will be fed into this network for obtaining the hidden feature distribution), where the formula is shown below:
\begin{equation}
    FID = \lVert m-m_w\rVert^2+Trace (c+c_w-2(cc_w)^{\frac{1}{2}})
\end{equation}
where the $m$ and $m_w$ represent the mean values of the latent features from the generated motions and real motions, respectively. while $c$ and $c_w$ denote the corresponding covariance matrices. The latent features we use are from the L3 hidden layer of architecture~\cite{holden2015learning}. For more detailed information about the architecture, please refer to Appendix A.

The comparison results are shown in Table~\ref{tab:fid} and a smaller score indicates better performance. We compare our method with vanilla SAN~\cite{aberman2020skeleton} as well as NKN$^*$ and  PMnet$^*$ (see Sec~\ref{sec:mixamo_experiment} for their definitions). We denote the evaluation item as $Biped\rightarrow Quadruped$ since we retarget the bipedal motions to the quadruped and compute the distribution with the real quadrupedal motions. $Quadruped\rightarrow Biped$ can be calculated in the opposite direction. To make the calculation of the metric insensitive to the root velocity distribution, 
% (assuming that in the extreme case the source motion dataset contains only stationary poses and thus the retargeted motions are also stationary, while the real motion dataset has locomotions)
we uniformly sample 1200 motion clips (64 frames each) in each \red{test set} based on the root velocity distribution \red{to represent the real motion distribution}. The results shown in Table~\ref{tab:fid} illustrate that we outperform other competing methods thanks to our advanced spatial modeling and body-part corresponding strategy. \HL{SAN fails on this task because the convolution operation based on the skeleton neighborhood leads to an unreasonable semantic correspondence between these two skeletons. NKN$^*$ and PMnet$^*$ do not consider any spatial relationship, resulting in inferior performance to our method.} 

\begin{figure}[ht]
    \centering
    \includegraphics[width=\linewidth]{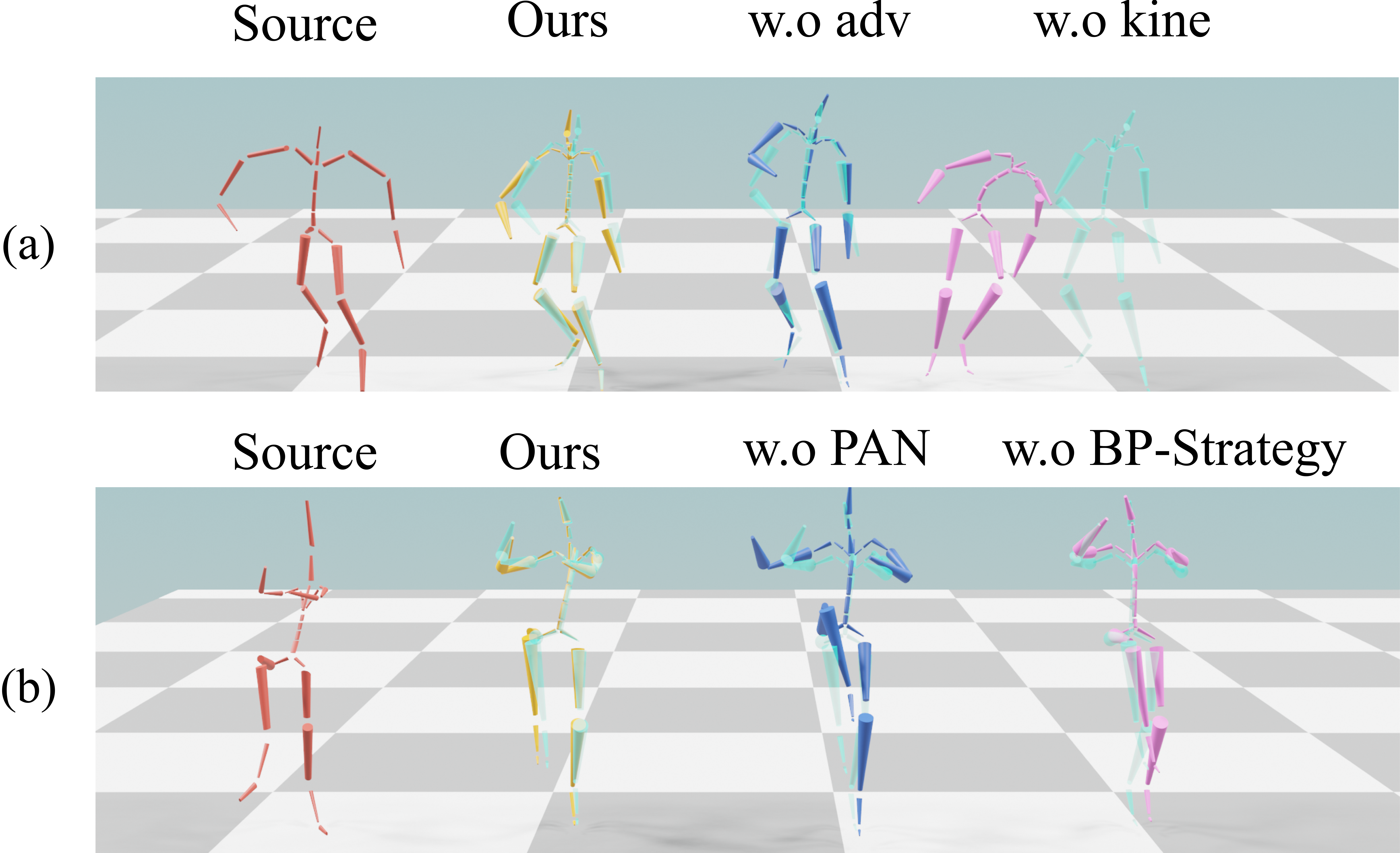}
    \caption{Qualitative results of ablation study. We remove some modules from the architecture to compare with our full method. The outputs are overlaid with the ground truth(transparent cyan skeleton).} 
    \label{fig:ablation_study}
\end{figure}
\subsubsection{User Study}
FID is capable of depicting the naturalness of the retargeted motions, but the similarity of retargeted motion to the source is difficult to quantify in the absence of ground truth. Therefore, we choose to judge the similarity by a user study. For each direction of retargeting($Quadruped\rightarrow Biped, Biped\rightarrow Quadruped$) we choose three types of action: $idle, move,$ and $sit$. The participants are asked to grade the similarity between the retargeted and source motions\redd{ like Sec~\ref{sec:h36m}} We run this user study on a total of 20 users and 12 questions, each is a random sample from the datasets. The action types are labeled automatically, similar to~\cite{2020Local}, i.e. we determine whether the action type is $move$ or $idle$ by the root velocity magnitude. As for the $sit$ label, we detect whether the hip/tail joint touches the ground during the motion. Table~\ref{tab:user_study} show the mean and standard deviation of the performance scores of each method on different actions. The scores show that we outperform other competing methods on most action types, which demonstrates our method also achieves more visually plausible retargeting from a subjective perspective.

\section{Ablation Study}\label{sec:ablation}
We conduct an ablation study to further demonstrate the contribution of each component in our architecture and help us understand the design of our
framework. We provide several examples under different self-comparison settings in Figure~\ref{fig:ablation_study} and the quantitative results are detailed in Table~\ref{tab:mse_ablation}. The evaluation also can be found in the supplementary video.

\begin{table}[ht]
     \caption{Quantitative Results of Ablation Study. We Remove Some Components of Our Architecture or Loss Terms from Our Full Method to Evaluate the Retargeting Performance on the Mixamo Dataset.}
    \centering
    \begin{tabular}{c||c||c}
        \hline
         ~ & Intra-Structural  &  Cross-Structural \\
        \hline\hline
         ours & \textbf{0.50} & \textbf{1.62} \\
         w.o BP-strategy & 0.58 & 2.29 \\ 
         w.o PAN & 1.19& 2.28 \\ 
         w.o $\mathcal{L}_{kine}$ & 1.11 & 48.08 \\
         w.o $\mathcal{L}_{adv}$ & 1.16 & 3.23\\
    \end{tabular}
    \label{tab:mse_ablation}
\end{table}

\subsection{Effect of Pose-Aware Attention}
To demonstrate the effect of the pose-aware attention mechanism in our architecture, we compare our full approach with the model without the attention mechanism. Specifically, we remove the attention operations described by equation~\ref{eq:QKV} and \ref{eq:attention_function}, replace them with a simple MLP layer, and keep the rest of our architecture fixed. When the pose-aware attention is removed, the model can not dynamically process the spatial features. The results shown in Figure~\ref{fig:ablation_study} (b) and Table~\ref{tab:mse_ablation} both demonstrate that the pose-aware attention mechanism can improve the accuracy of the retargeting. In addition, the recall curves of foot contact in Figure~\ref{fig:contact_recall} also show that the proposed attention mechanism is beneficial to improve the stability of the retargeting. 

\subsection{Effect of Body Part Strategy}
To illustrate the effectiveness of the body part strategy (BP-strategy), we replace the "body part tokens" with "whole body token", i.e., we remove the mask matrix $U$ from equation~\ref{eq:attention_function} so that all joints are associated with the only "whole body token". In addition, the convolution kernels in equation~\ref{eq:conv} are not body-part related and will degrade to the vanilla 1D convolution kernels. The results in Table~\ref{tab:mse_ablation} and Figure~\ref{fig:foot_sliding} both show some decrease in performance for the model without the BP-strategy. We believe that the body parts provide a geometric prior for the neural networks and using body parts as the retargeting units is beneficial for network convergence when an unsupervised training manner is used.

\subsection{Effect of Kinematic Loss}
We investigate the effectiveness of kinematic loss by removing the loss term $\mathcal{L}_{kine}$ during training. The kinematic loss can supervise the training in Cartesian space which is more crucial to our perception. From Table~\ref{tab:mse_ablation}, we observe that our full model performs better than the model without this loss, especially in the Cross-Structural retargeting since the Cross-Structural retargeting is harder than the Intra-Structural retargeting. We believe that the rotation is more sensitive than the joint position since the rotation error of the father joint will be amplified in the coordinate positions of the child's joint by matrix chain multiplication. Therefore, we must introduce the kinematic loss when training the networks. 

\subsection{Effect of Adversarial Loss}
To evaluate the contribution of adversarial loss, we discard the loss term $\mathcal{L}_{adv}$ and retrain our networks. The results in Table~\ref{tab:mse_ablation} show that our full method outperforms the model without adversarial loss term in both retargeting scenarios. We believe that $\mathcal{L}_{adv}$ is very important in unsupervised learning since it ensures that the motion generated by the networks falls in the motion manifold of the corresponding structure. The example shown in Figure~\ref{fig:ablation_study} (a) also illustrates the importance of the adversarial loss term in our retargeting framework.
\section{Discussion and Future Work}
We propose a novel motion retargeting framework that uses body parts as the basic retargeting units, which together with our pose-aware attention mechanism can dynamically extract the spatial features of the motion. Our architecture can learn the shared motion space of body parts from unstructured motion capture data, which can easily allow retargeting among skeletons with different structure. Due to our dynamical spatial modeling, our method allows for more accurate and flexible retargeting compared to other approaches. In particular, we significantly improve the performance of motion retargeting between quadrupeds and bipeds. 

\redd{The main} limitation of our work is the dependence on motion statistics, i.e., we still require an amount of \redd{balanced} motion capture data for each structure. \redd{Both datasets need to contain motions with various facing directions, velocity directions, velocity intervals, and the same action types as much as possible. We show in the supplemental video a failure case when retargeting the backward walking of a biped to a quadruped, where we find that the quadruped moves unnaturally and without gaits. This is because, although the biped training set includes backward walking, the quadruped training set lacks such movement, making it difficult to correspond when learning. When we test the retargeting of bipedal artistic movements, such as hopping (unseen action) to the quadruped. We find that our method can only use some corresponding laws learned from locomotion to generate the retargeted motion of the quadruped, which cannot guarantee naturalness and rhythm. This motivates us to introduce more control signals to achieve anthropomorphic retargeting in the future.} 

In some scenarios, we do not have access to the capture data of the target skeleton. For this problem, we can consider introducing one-shot or zero-shot learning into motion retargeting in the future, which can further expand the application of automatic motion retargeting. \red{Another drawback of our work is that we currently rely on self-supervised and adversarial learning to align the motion manifolds of different structure. This \redd{correspondence} may result in some output being semantically inconsistent with the source motion and lacking physical realism, \redd{e.g., distinguishing between sitting on the floor and sitting on a chair when retargeting the biped motion to the quadruped. }}\red{ One potential direction is to combine our spatial modeling with the Dynamic Motion Reassembly from Virtual Chimeras~\cite{lee2022learning} to achieve physically realistic motion retargeting.}

From another perspective, our approach is in fact to learn the motion manifolds of various body parts. Therefore, we are also able to implement motion editing or user interaction in the latent space like~\cite{holden2015learning, villegas2021contact} \red{to achieve body part level control in the future}. \red{Recently, DeepPhase~\cite{starke2022deepphase} starts to learn motion manifold from a temporal alignment perspective, which can extract periodic features of the whole body. This alignment is very helpful for retargeting between different organisms because it is difficult for us to obtain pairwise motion data. We are able to introduce periodicity constraints into the loss function to make the generated motions more reasonable.}

% if have a single appendix:
%\appendix[Proof of the Zonklar Equations]
% or
%\appendix  % for no appendix heading
% do not use \section anymore after \appendix, only \section*
% is possibly needed

% use appendices with more than one appendix
% then use \section to start each appendix
% you must declare a \section before using any
% \subsection or using \label (\appendices by itself
% starts a section numbered zero.)
%

\appendices
% \section{The architecture of the Autoencoder For FID Calculation}\label{app:A}
% The network architecture we use for FID calculation is described in Table~\ref{tab:fid_archi}. In this table, s and k are short for the stride size and kernel size, respectively. up is meaning the upsample scale, we use the linear mode and set align\_corners = Flase in implementation. The value in parentheses after LRelu indicates the negative slope and the values in parentheses after MaxPool represent kernel size and stride, respectively. 

% use section* for acknowledgment
\ifCLASSOPTIONcompsoc
  % The Computer Society usually uses the plural form
  \section*{Acknowledgments}
\else
  % regular IEEE prefers the singular form
  \section*{Acknowledgment}
\fi

This work was supported by National Key R\&D Program "Industrial Software" Special Project of China (NO. 2022YFB3303202).

% Can use something like this to put references on a page
% by themselves when using endfloat and the captionsoff option.
\ifCLASSOPTIONcaptionsoff
  \newpage
\fi

% trigger a \newpage just before the given reference
% number - used to balance the columns on the last page
% adjust value as needed - may need to be readjusted if
% the document is modified later
%\IEEEtriggeratref{8}
% The "triggered" command can be changed if desired:
%\IEEEtriggercmd{\enlargethispage{-5in}}

% references section

% can use a bibliography generated by BibTeX as a .bbl file
% BibTeX documentation can be easily obtained at:
% http://mirror.ctan.org/biblio/bibtex/contrib/doc/
% The IEEEtran BibTeX style support page is at:
% http://www.michaelshell.org/tex/ieeetran/bibtex/
%\bibliographystyle{IEEEtran}
% argument is your BibTeX string definitions and bibliography database(s)
%\bibliography{IEEEabrv,../bib/paper}
%
% <OR> manually copy in the resultant .bbl file
% set second argument of \begin to the number of references
% (used to reserve space for the reference number labels box)

\bibliographystyle{IEEEtran}
\bibliography{main}

% \begin{thebibliography}{1}

% \bibitem{IEEEhowto:kopka}
% H.~Kopka and P.~W. Daly, \emph{A Guide to \LaTeX}, 3rd~ed.\hskip 1em plus
%   0.5em minus 0.4em\relax Harlow, England: Addison-Wesley, 1999.

% \end{thebibliography}

% biography section
% 
% If you have an EPS/PDF photo (graphicx package needed) extra braces are
% needed around the contents of the optional argument to biography to prevent
% the LaTeX parser from getting confused when it sees the complicated
% \includegraphics command within an optional argument. (You could create
% your own custom macro containing the \includegraphics command to make things
% simpler here.)
\vspace{-20pt}
\begin{IEEEbiography}[{\includegraphics[width=1in,height=1.25in,clip,keepaspectratio]{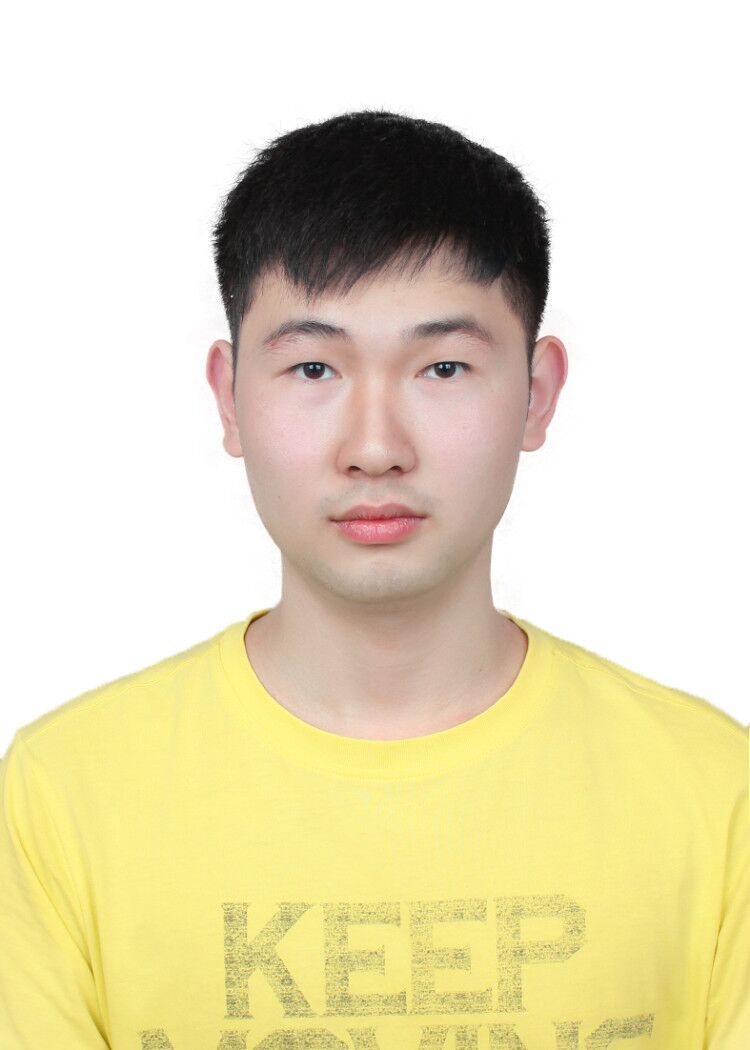}}]{Lei Hu}
% or if you just want to reserve a space for a photo:
Lei Hu received a B.Sc. degree in mathematics and applied mathematics from Southwest Jiaotong University (SWJTU), China, in 2019. He is currently pursuing a Ph.D. degree in computer science at the Institute of Computing Technology, Chinese Academy of Sciences, supervised by Prof. Shihong Xia.
\end{IEEEbiography}
\vspace{-45pt}
\begin{IEEEbiography}[{\includegraphics[width=1in,height=1.25in,clip,keepaspectratio]{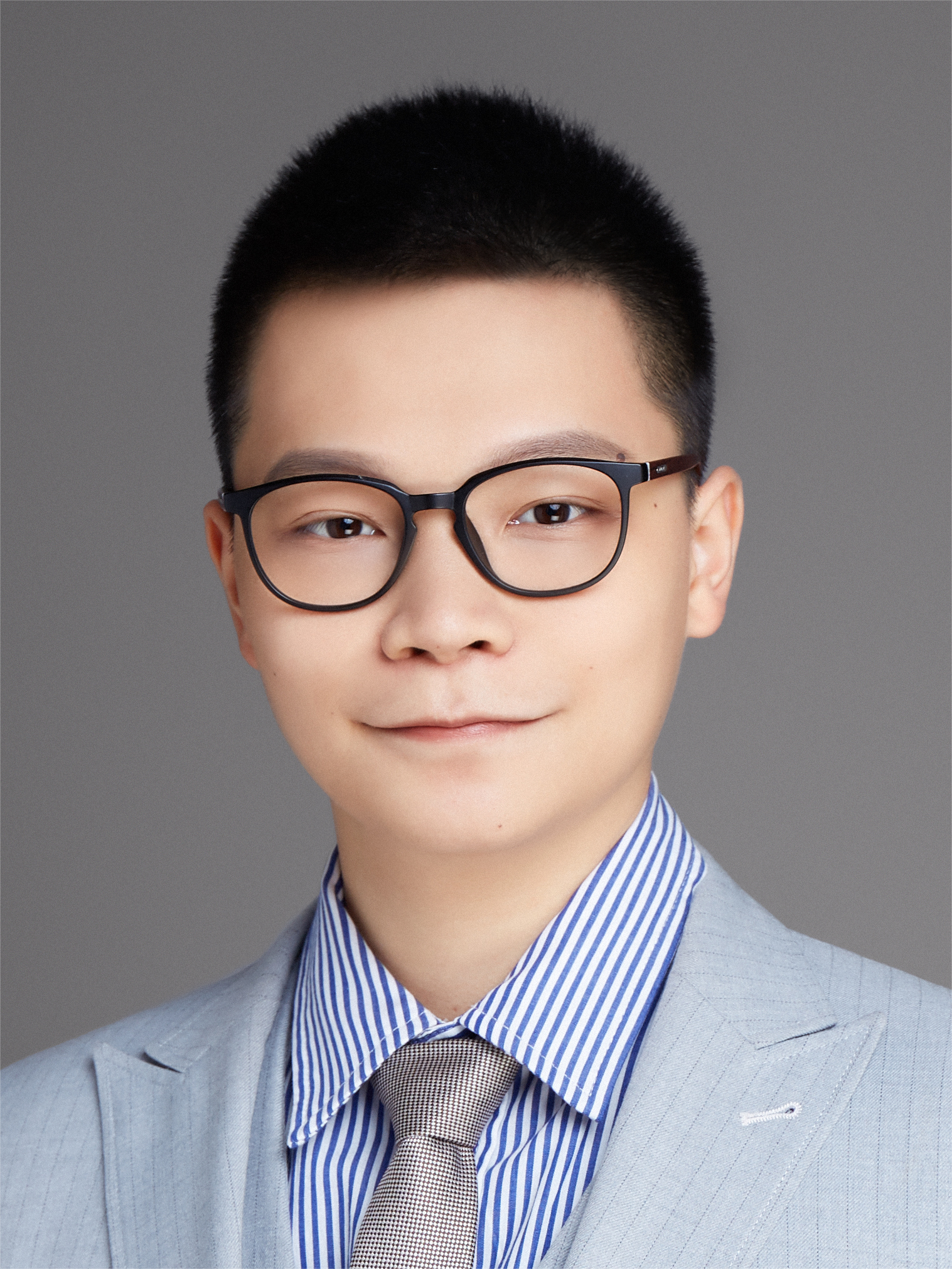}}]{Zihao Zhang}
Zihao Zhang (Member, IEEE) received a B.Sc. degree in mathematics from Sichuan University, Sichuan, China, in 2016, and a Ph.D. degree from the Institute of Computing Technology, Chinese Academy of Sciences, Beijing, China, in 2022.  He is currently a special research assistant with the Intelligent Processor Research Center at the Institute of Computing Technology, Chinese Academy of Sciences. His research interests include human motion modeling and image restoration.
\end{IEEEbiography}
\vspace{-45pt}
\begin{IEEEbiography}[{\includegraphics[width=1in,height=1.25in,clip,keepaspectratio]{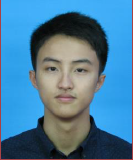}}]{Chongyang Zhong}
Chongyang Zhong received a B.Sc. degree in automation from Tsinghua University (THU), China, in 2017. He is currently pursuing a Ph.D. degree in computer science at the Institute of Computing Technology, Chinese Academy of Sciences, supervised by Prof. Shihong Xia.
\end{IEEEbiography}
\vspace{-45pt}
\begin{IEEEbiography}[{\includegraphics[width=1in,height=1.25in,clip,keepaspectratio]{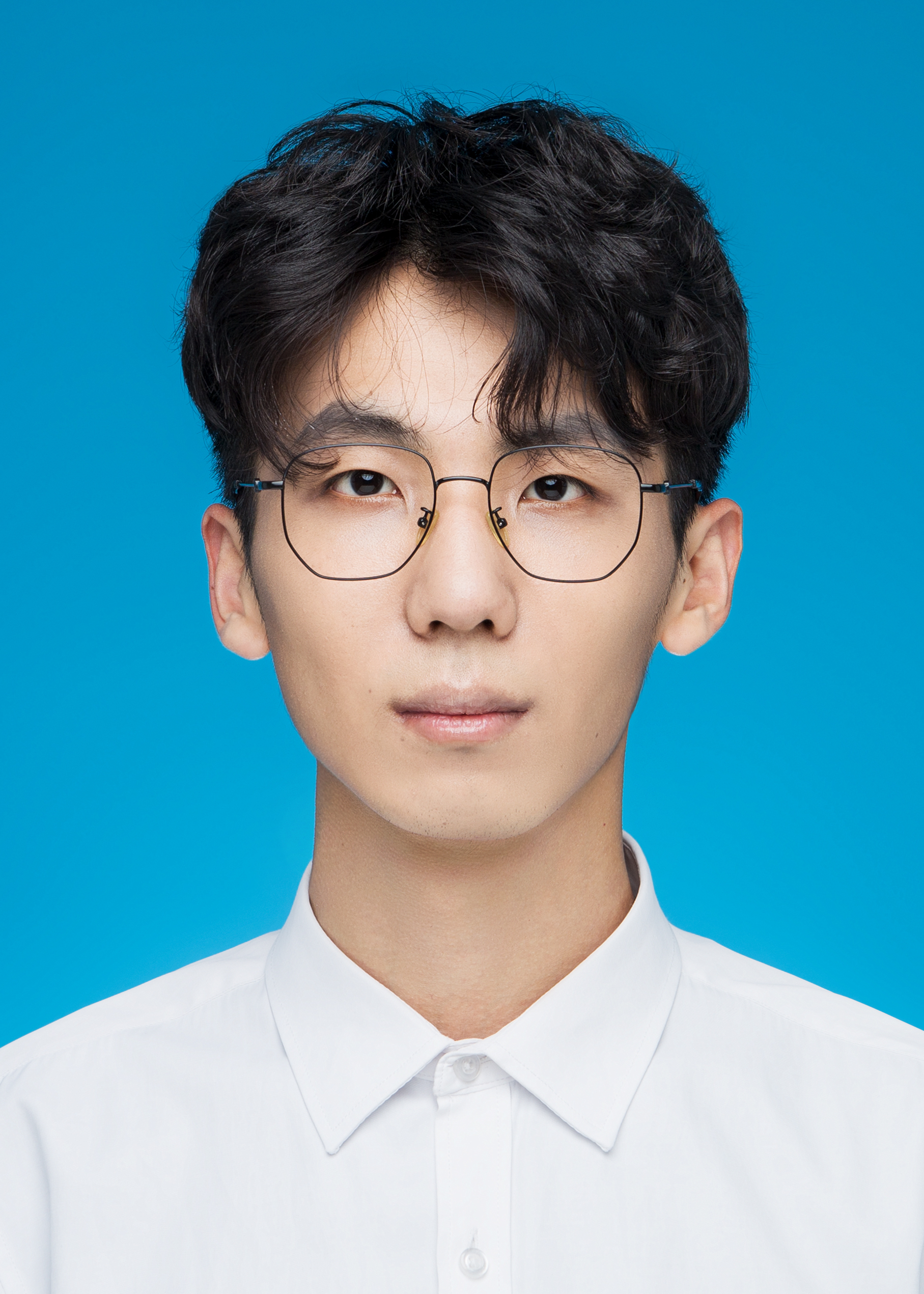}}]{Boyuan Jiang}
Boyuan Jiang received a B.Sc. degree in mathematics and applied mathematics from China University of Geosciences, Beijing(CUGB), China, in 2020. He is currently pursuing an M.E. degree in computer science at the Institute of Computing Technology, Chinese Academy of Sciences, supervised by Prof. Shihong Xia.
\end{IEEEbiography}
% if you will not have a photo at all:
\vspace{-45pt}
\begin{IEEEbiography}[{\includegraphics[width=1in,height=1.25in,clip,keepaspectratio]{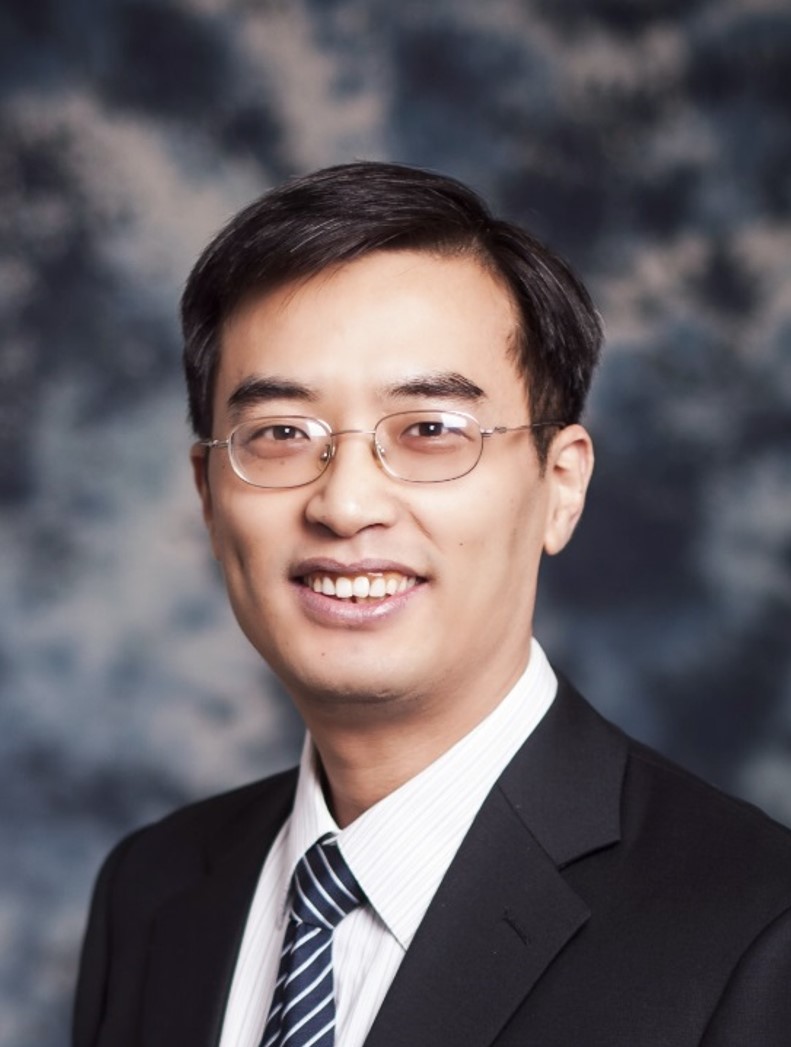}}]{Shihong Xia}
Shihong Xia received a Ph.D. degree in computer science from the University of Chinese Academy of Sciences. He is currently a professor at the Institute of Computing Technology, Chinese Academy of Sciences (ICT, CAS), and the director of the human motion modeling laboratory. His primary research is in the area of computer
graphics, virtual reality, and artificial intelligence.
\end{IEEEbiography}

% insert where needed to balance the two columns on the last page with
% biographies
%\newpage

% You can push biographies down or up by placing
% a \vfill before or after them. The appropriate
% use of \vfill depends on what kind of text is
% on the last page and whether or not the columns
% are being equalized.

%\vfill

% Can be used to pull up biographies so that the bottom of the last one
% is flush with the other column.
%\enlargethispage{-5in}

% that's all folks
\end{document}

\markboth{Journal of \LaTeX\ Class Files}
{Shell \MakeLowercase{\textit{et al.}}: Bare Demo of IEEEtran.cls for Computer Society Journals}
\IEEEdisplaynontitleabstractindextext

\appendices
\section{The architecture of the Autoencoder For FID Calculation}\label{app:A}
The network architecture we use for FID calculation is described in Table~\ref{tab:fid_archi}. In this table, s and k are short for the stride size and kernel size, respectively. up is meaning the upsample scale, we use the linear mode and set align\_corners = Flase in implementation. The value in parentheses after LRelu indicates the negative slope and the values in parentheses after MaxPool represent kernel size and stride, respectively. 

\begin{table}[ht]
    \caption{Architecture of the Autoencoder for FID Calculation.}
    \centering
    \begin{tabular}{ccccc}
        \hline
         Layer & Size & Filter & Act. & Pooling\\
        \hline
         Conv1 & 256 & k=15,s=2 & LReLU(0.2) & MaxPool(2, 2) \\
         Conv2  & 256 & k=15,s=2 & LReLU(0.2) & MaxPool(2, 2) \\
         Conv3 & 256 & k=3,s=2 & Tanh & - \\
         Deconv1 & 256 & k=3,up=2 & LReLU(0.2) & - \\
         Deconv2 & 256 & k=15,up=4 & LReLU(0.2) & - \\
         Deconv3 & 256 & k=15,up=4 & - & -  \\
        \hline
    \end{tabular}
    \label{tab:fid_archi}
\end{table}

\section{The train-test split for retargeting between quadrupeds and bipeds }\label{app:B}

For the bipedal dataset LaFan1, we use only four motion types: aiming, walking, running, and sprinting for training and testing. As shown in Table~\ref{tab:dataset_split}, there are 5 subjects, we used subject 1 for testing and the rest for training. In the quadrupedal dataset, we split the whole 52 motion files randomly and make sure the train-test ratio is close to the bipedal dataset split, which is shown on the right side of Table~\ref{tab:dataset_split}

\begin{table}[ht]
    \caption{Train-Test Set Split for Biped-To-Quadruped Setting.}
    \centering
    \begin{tabular}{cc|cc}
        \hline
        \multicolumn{2}{c|}{Biped} & \multicolumn{2}{c}{Quadruped}\\
        \hline
         Train & Test & Train & Test \\
        \hline
         aiming1\_subject4 & aiming1\_subject1 & \multirow{17}{*}{\makecell{All motion files\\ except for those \\appearing in the \\test set}} &D1\_001\_KAN01\_001\\
         aiming2\_subject2 & walk1\_subject1 & ~ &D1\_004\_KAN01\_001  \\
         aiming2\_subject3 & walk2\_subject1 &~ &D1\_005\_KAN01\_001 \\
         aiming2\_subject5 & walk3\_subject1 & ~ &D1\_008\_KAN01\_002  \\
         run1\_subject2 & walk4\_subject1 &~ &D1\_010\_KAN01\_001  \\
         run1\_subject5 & run2\_subject1 &~ &D1\_013\_KAN01\_001  \\
         run2\_subject4 & ~ &~ &D1\_047\_KAN01\_001  \\
         sprint1\_subject2 & ~ &~ &D1\_047z\_KAN01\_002  \\
         sprint1\_subject4 & ~ &~ &D1\_047z\_KAN01\_003  \\
         walk1\_subject2 & ~ &~ &D1\_057\_KAN01\_001  \\
         walk1\_subject5 & ~ &~ &D1\_061z\_KAN01\_003  \\
         walk2\_subject3 & ~ &~ &D1\_ex03\_KAN02\_006  \\
         walk2\_subject4 & ~ &~ &D1\_ex04\_KAN02\_001  \\
         walk3\_subject2 & ~ &~ &~  \\
         walk3\_subject3 & ~ &~ &~  \\
         walk3\_subject4 & ~ &~ &~  \\
         walk3\_subject5 & ~ &~ &~  \\

        \hline
    \end{tabular}
    \label{tab:dataset_split}
\end{table}

% you can choose not to have a title for an appendix
% if you want by leaving the argument blank

% use section* for acknowledgment
% \ifCLASSOPTIONcompsoc
%   % The Computer Society usually uses the plural form
%   \section*{Acknowledgments}
% \else
%   % regular IEEE prefers the singular form
%   \section*{Acknowledgment}
% \fi

% The authors would like to thank...

% Can use something like this to put references on a page
% by themselves when using endfloat and the captionsoff option.
% \ifCLASSOPTIONcaptionsoff
%   \newpage
% \fi

% \bibliographystyle{IEEEtran}
% \bibliography{mybib}